\documentstyle[12pt]{article}
\def\fileversion{v1.13}%
\def\filedate{6.5.93}%
\edef\epsfigRestoreAt{\catcode`@=\number\catcode`@\relax}%
\catcode`\@=11\relax
\immediate\write16{Document style option `epsfig', \fileversion\space
<\filedate> (edited by SPQR)}%
\newcount\EPS@Height
\newcount\EPS@Width
\newcount\EPS@xscale
\newcount\EPS@yscale
\def\psfigdriver#1{%
  \bgroup\edef\next{\def\noexpand\tempa{#1}}%
    \uppercase\expandafter{\next}%
    \def\LN{DVITOLN03}%
    \def\DVItoPS{DVITOPS}%
    \def\DVIPS{DVIPS}%
    \def\emTeX{EMTEX}%
    \def\OzTeX{OZTEX}%
    \def\Textures{TEXTURES}%
    \global\chardef\fig@driver=0
    \ifx\tempa\LN
        \global\chardef\fig@driver=0\fi
    \ifx\tempa\DVItoPS
        \global\chardef\fig@driver=1\fi
    \ifx\tempa\DVIPS
        \global\chardef\fig@driver=2\fi
    \ifx\tempa\emTeX
        \global\chardef\fig@driver=3\fi
    \ifx\tempa\OzTeX
        \global\chardef\fig@driver=4\fi
    \ifx\tempa\Textures
        \global\chardef\fig@driver=5\fi
  \egroup
\def\psfig@start{}%
\def\psfig@end{}%
\def\epsfig@gofer{}%
\ifcase\fig@driver
\typeout{WARNING! ****
 no specials for LN03 psfig}%
\or 
\def\psfig@start{}%
\def\psfig@end{\special{dvitops: import \@p@sfilefinal \space
\@p@swidth sp \space \@p@sheight sp \space fill}%
\if@clip \typeout{Clipping not supported}\fi
\if@angle \typeout{Rotating not supported}\fi
}%
\let\epsfig@gofer\psfig@end
\or 
\def\psfig@start{\special{ps::[begin]  \@p@swidth \space \@p@sheight \space%
        \@p@sbbllx \space \@p@sbblly \space%
        \@p@sbburx \space \@p@sbbury \space%
        startTexFig \space }%
        \if@angle
                \special {ps:: \@p@sangle \space rotate \space}
        \fi
        \if@clip
                \if@verbose
                        \typeout{(clipped to BB) }%
                \fi
                \special{ps:: doclip \space }%
        \fi
        \special{ps: plotfile \@p@sfilefinal \space }%
        \special{ps::[end] endTexFig \space }%
}%
\def\psfig@end{}%
\def\epsfig@gofer{\if@clip
                        \if@verbose
                           \typeout{(clipped to BB)}%
                        \fi
                        \epsfclipon
                  \fi
                  \epsfsetgraph{\@p@sfilefinal}%
}%
\or 
\typeout{WARNING. You must have a .bb info file with the Bounding Box
  of the pcx file}%
\def\psfig@start{}%
\def\psfig@end{\typeout{pcx import of \@p@sfilefinal}%
\if@clip \typeout{Clipping not supported}\fi
\if@angle \typeout{Rotating not supported}\fi
\raisebox{\@p@srheight true sp}{\special{em: graph \@p@sfilefinal}}}%
\def\epsfig@gofer{}%
\or 
\def\psfig@start{}%
\def\psfig@end{%
\EPS@Width\@p@swidth
\EPS@Height\@p@sheight
\divide\EPS@Width by 65781  
\divide\EPS@Height by 65781
\special{epsf=\@p@sfilefinal
\space
width=\the\EPS@Width
\space
height=\the\EPS@Height
}%
\if@clip \typeout{Clipping not supported}\fi
\if@angle \typeout{Rotating not supported}\fi
}%
\let\epsfig@gofer\psfig@end
\or 
\def\psfig@end{\if@clip
                        \if@verbose
                           \typeout{(clipped to BB)}%
                        \fi
                        \epsfclipon
                  \fi
\special{illustration \@p@sfilefinal\space scaled \the\EPS@xscale}%
}%
\def\psfig@start{}%
\let\epsfig\psfig
\else
\typeout{WARNING. *** unknown  driver - no psfig}%
\fi
}%
\newdimen\ps@dimcent
\ifx\undefined\fbox
\newdimen\fboxrule
\newdimen\fboxsep
\newdimen\ps@tempdima
\newbox\ps@tempboxa
\long\def\fbox#1{\leavevmode\setbox\ps@tempboxa\hbox{#1}\ps@tempdima\fboxrule
    \advance\ps@tempdima \fboxsep \advance\ps@tempdima \dp\ps@tempboxa
   \hbox{\lower \ps@tempdima\hbox
  {\vbox{\hrule height \fboxrule
          \hbox{\vrule width \fboxrule \hskip\fboxsep
          \vbox{\vskip\fboxsep \box\ps@tempboxa\vskip\fboxsep}\hskip
                 \fboxsep\vrule width \fboxrule}%
                 \hrule height \fboxrule}}}}%
\fi
\ifx\@ifundefined\undefined
\long\def\@ifundefined#1#2#3{\expandafter\ifx\csname
  #1\endcsname\relax#2\else#3\fi}%
\fi
\@ifundefined{typeout}%
{\gdef\typeout#1{\immediate\write\sixt@@n{#1}}}%
{\relax}%
\@ifundefined{epsfig}{}{\typeout{EPSFIG --- already loaded}\endinput}%
\@ifundefined{epsfbox}{\input epsf}{}%
\ifx\undefined\@latexerr
        \newlinechar`\^^J
        \def\@spaces{\space\space\space\space}%
        \def\@latexerr#1#2{%
        \edef\@tempc{#2}\expandafter\errhelp\expandafter{\@tempc}%
        \typeout{Error. \space see a manual for explanation.^^J
         \space\@spaces\@spaces\@spaces Type \space H <return> \space for
         immediate help.}\errmessage{#1}}%
\fi
\def\@whattodo{You tried to include a PostScript figure which
cannot be found^^JIf you press return to carry on anyway,^^J
The failed name will be printed in place of the figure.^^J
or type X to quit}%
\def\@whattodobb{You tried to include a PostScript figure which
has no^^Jbounding box, and you supplied none.^^J
If you press return to carry on anyway,^^J
The failed name will be printed in place of the figure.^^J
or type X to quit}%
\def\@nnil{\@nil}%
\def\@empty{}%
\def\@psdonoop#1\@@#2#3{}%
\def\@psdo#1:=#2\do#3{\edef\@psdotmp{#2}\ifx\@psdotmp\@empty \else
    \expandafter\@psdoloop#2,\@nil,\@nil\@@#1{#3}\fi}%
\def\@psdoloop#1,#2,#3\@@#4#5{\def#4{#1}\ifx #4\@nnil \else
       #5\def#4{#2}\ifx #4\@nnil \else#5\@ipsdoloop #3\@@#4{#5}\fi\fi}%
\def\@ipsdoloop#1,#2\@@#3#4{\def#3{#1}\ifx #3\@nnil
       \let\@nextwhile=\@psdonoop \else
      #4\relax\let\@nextwhile=\@ipsdoloop\fi\@nextwhile#2\@@#3{#4}}%
\def\@tpsdo#1:=#2\do#3{\xdef\@psdotmp{#2}\ifx\@psdotmp\@empty \else
    \@tpsdoloop#2\@nil\@nil\@@#1{#3}\fi}%
\def\@tpsdoloop#1#2\@@#3#4{\def#3{#1}\ifx #3\@nnil
       \let\@nextwhile=\@psdonoop \else
      #4\relax\let\@nextwhile=\@tpsdoloop\fi\@nextwhile#2\@@#3{#4}}%
\long\def\epsfaux#1#2:#3\\{\ifx#1\epsfpercent
   \def\testit{#2}\ifx\testit\epsfbblit
        \@atendfalse
        \epsf@atend #3 . \\%
        \if@atend
           \if@verbose
                \typeout{epsfig: found `(atend)'; continuing search}%
           \fi
        \else
                \epsfgrab #3 . . . \\%
                \epsffileokfalse\global\no@bbfalse
                \global\epsfbbfoundtrue
        \fi
   \fi\fi}%
\def\epsf@atendlit{(atend)}
\def\epsf@atend #1 #2 #3\\{%
   \def\epsf@tmp{#1}\ifx\epsf@tmp\empty
      \epsf@atend #2 #3 .\\\else
   \ifx\epsf@tmp\epsf@atendlit\@atendtrue\fi\fi}%

\chardef\trig@letter = 11
\chardef\other = 12

\newif\ifdebug 
\newif\ifc@mpute 
\newif\if@atend
\c@mputetrue 

\let\then = \relax
\def\r@dian{pt }%
\let\r@dians = \r@dian
\let\dimensionless@nit = \r@dian
\let\dimensionless@nits = \dimensionless@nit
\def\internal@nit{sp }%
\let\internal@nits = \internal@nit
\newif\ifstillc@nverging
\def \Mess@ge #1{\ifdebug \then \message {#1} \fi}%

{ 
        \catcode `\@ = \trig@letter
        \gdef \nodimen {\expandafter \n@dimen \the \dimen}%
        \gdef \term #1 #2 #3%
               {\edef \t@ {\the #1}
                \edef \t@@ {\expandafter \n@dimen \the #2\r@dian}%
                \t@rm {\t@} {\t@@} {#3}%
               }%
        \gdef \t@rm #1 #2 #3%
               {{%
                \count 0 = 0
                \dimen 0 = 1 \dimensionless@nit
                \dimen 2 = #2\relax
                \Mess@ge {Calculating term #1 of \nodimen 2}%
                \loop
                \ifnum  \count 0 < #1
                \then   \advance \count 0 by 1
                        \Mess@ge {Iteration \the \count 0 \space}%
                        \Multiply \dimen 0 by {\dimen 2}%
                        \Mess@ge {After multiplication, term = \nodimen 0}%
                        \Divide \dimen 0 by {\count 0}%
                        \Mess@ge {After division, term = \nodimen 0}%
                \repeat
                \Mess@ge {Final value for term #1 of
                                \nodimen 2 \space is \nodimen 0}%
                \xdef \Term {#3 = \nodimen 0 \r@dians}%
                \aftergroup \Term
               }}%
        \catcode `\p = \other
        \catcode `\t = \other
        \gdef \n@dimen #1pt{#1} 
}%

\def \Divide #1by #2{\divide #1 by #2} 

\def \Multiply #1by #2
       {{
        \count 0 = #1\relax
        \count 2 = #2\relax
        \count 4 = 65536
        \Mess@ge {Before scaling, count 0 = \the \count 0 \space and
                        count 2 = \the \count 2}%
        \ifnum  \count 0 > 32767 
        \then   \divide \count 0 by 4
                \divide \count 4 by 4
        \else   \ifnum  \count 0 < -32767
                \then   \divide \count 0 by 4
                        \divide \count 4 by 4
                \else
                \fi
        \fi
        \ifnum  \count 2 > 32767 
        \then   \divide \count 2 by 4
                \divide \count 4 by 4
        \else   \ifnum  \count 2 < -32767
                \then   \divide \count 2 by 4
                        \divide \count 4 by 4
                \else
                \fi
        \fi
        \multiply \count 0 by \count 2
        \divide \count 0 by \count 4
        \xdef \product {#1 = \the \count 0 \internal@nits}%
        \aftergroup \product
       }}%

\def\r@duce{\ifdim\dimen0 > 90\r@dian \then   
                \multiply\dimen0 by -1
                \advance\dimen0 by 180\r@dian
                \r@duce
            \else \ifdim\dimen0 < -90\r@dian \then  
                \advance\dimen0 by 360\r@dian
                \r@duce
                \fi
            \fi}%

\def\Sine#1%
       {{%
        \dimen 0 = #1 \r@dian
        \r@duce
        \ifdim\dimen0 = -90\r@dian \then
           \dimen4 = -1\r@dian
           \c@mputefalse
        \fi
        \ifdim\dimen0 = 90\r@dian \then
           \dimen4 = 1\r@dian
           \c@mputefalse
        \fi
        \ifdim\dimen0 = 0\r@dian \then
           \dimen4 = 0\r@dian
           \c@mputefalse
        \fi
        \ifc@mpute \then
                \divide\dimen0 by 180
                \dimen0=3.141592654\dimen0
                \dimen 2 = 3.1415926535897963\r@dian 
                \divide\dimen 2 by 2 
                \Mess@ge {Sin: calculating Sin of \nodimen 0}%
                \count 0 = 1 
                \dimen 2 = 1 \r@dian 
                \dimen 4 = 0 \r@dian 
                \loop
                        \ifnum  \dimen 2 = 0 
                        \then   \stillc@nvergingfalse
                        \else   \stillc@nvergingtrue
                        \fi
                        \ifstillc@nverging 
                        \then   \term {\count 0} {\dimen 0} {\dimen 2}%
                                \advance \count 0 by 2
                                \count 2 = \count 0
                                \divide \count 2 by 2
                                \ifodd  \count 2 
                                \then   \advance \dimen 4 by \dimen 2
                                \else   \advance \dimen 4 by -\dimen 2
                                \fi
                \repeat
        \fi
                        \xdef \sine {\nodimen 4}%
       }}%

\def\Cosine#1{\ifx\sine\UnDefined\edef\Savesine{\relax}\else
                             \edef\Savesine{\sine}\fi
        {\dimen0=#1\r@dian\multiply\dimen0 by -1
         \advance\dimen0 by 90\r@dian
         \Sine{\nodimen 0}%
         \xdef\cosine{\sine}%
         \xdef\sine{\Savesine}}}
\def\psdraft{\def\@psdraft{0}}%
\def\psfull{\def\@psdraft{1}}%
\psfull
\newif\if@scalefirst
\def\psscalefirst{\@scalefirsttrue}%
\def\psrotatefirst{\@scalefirstfalse}%
\psrotatefirst
\newif\if@draftbox
\def\psnodraftbox{\@draftboxfalse}%
\@draftboxtrue
\newif\if@noisy
\@noisyfalse
\newif\ifno@bb
\newif\if@bbllx
\newif\if@bblly
\newif\if@bburx
\newif\if@bbury
\newif\if@height
\newif\if@width
\newif\if@rheight
\newif\if@rwidth
\newif\if@angle
\newif\if@clip
\newif\if@verbose
\newif\if@prologfile
\def\@p@@sprolog#1{\@prologfiletrue\def\@prologfileval{#1}}%
\def\@p@@sclip#1{\@cliptrue}%
\newif\ifepsfig@dos  
\def\epsfigdos{\epsfig@dostrue}%
\epsfig@dosfalse
\newif\ifuse@psfig
\def\ParseName#1{\expandafter\@Parse#1}%
\def\@Parse#1.#2:{\gdef\BaseName{#1}\gdef\FileType{#2}}%

\def\@p@@sfile#1{%
\ifepsfig@dos
   \ParseName{#1:}%
\else
   \gdef\BaseName{#1}\gdef\FileType{}%
\fi
\def\@p@sfile{NO FILE: #1}%
\def\@p@sfilefinal{NO FILE: #1}%
        \openin1=#1
        \ifeof1\closein1
                \openin1=\BaseName.bb
                        \ifeof1\closein1
                                \if@bbllx\if@bblly\if@bburx\if@bbury
                                        \def\@p@sfile{#1}%
                                        \def\@p@sfilefinal{#1}%
                                        \fi\fi\fi
                                \else
                                        \@latexerr{ERROR.
PostScript file #1 not found}\@whattodo
                                        \@p@@sbbllx{100bp}%
                                        \@p@@sbblly{100bp}%
                                        \@p@@sbburx{200bp}%
                                        \@p@@sbbury{200bp}%
                                        \psdraft
                                \fi
                        \else
                                \closein1%
                                \edef\@p@sfile{\BaseName.bb}%
                                \typeout{using BB from \@p@sfile}%
                                \ifnum\fig@driver=3
                                  \edef\@p@sfilefinal{\BaseName.pcx}%
                                \else
                                 \ifepsfig@dos
                                 \edef\@p@sfilefinal{"`uncompress
                                   < \BaseName.Z"}%
                                \else
                                \edef\@p@sfilefinal{"`zcat `texfind
                                  #1.Z`"}%
                                \fi
                                \fi
                        \fi
        \else\closein1
                    \edef\@p@sfile{#1}%
                    \edef\@p@sfilefinal{#1}%
        \fi%
}%
\let\@p@@sfigure\@p@@sfile
\def\@p@@sbbllx#1{%
				            \@bbllxtrue
                \ps@dimcent=#1
                \edef\@p@sbbllx{\number\ps@dimcent}%
                \divide\ps@dimcent by65536
                \global\edef\epsfllx{\number\ps@dimcent}%
}%
\def\@p@@sbblly#1{%
                \@bbllytrue
                \ps@dimcent=#1
                \edef\@p@sbblly{\number\ps@dimcent}%
                \divide\ps@dimcent by65536
                \global\edef\epsflly{\number\ps@dimcent}%
}%
\def\@p@@sbburx#1{%
                \@bburxtrue
                \ps@dimcent=#1
                \edef\@p@sbburx{\number\ps@dimcent}%
                \divide\ps@dimcent by65536
                \global\edef\epsfurx{\number\ps@dimcent}%
}%
\def\@p@@sbbury#1{%
                \@bburytrue
                \ps@dimcent=#1
                \edef\@p@sbbury{\number\ps@dimcent}%
                \divide\ps@dimcent by65536
                \global\edef\epsfury{\number\ps@dimcent}%
}%
\def\@p@@sheight#1{%
                \@heighttrue
                \global\epsfysize=#1
                \ps@dimcent=#1
                \edef\@p@sheight{\number\ps@dimcent}%
}%
\def\@p@@swidth#1{%
                \@widthtrue
                \global\epsfxsize=#1
                \ps@dimcent=#1
                \edef\@p@swidth{\number\ps@dimcent}%
}%
\def\@p@@srheight#1{%
                \@rheighttrue\use@psfigtrue
                \ps@dimcent=#1
                \edef\@p@srheight{\number\ps@dimcent}%
}%
\def\@p@@srwidth#1{%
                \@rwidthtrue\use@psfigtrue
                \ps@dimcent=#1
                \edef\@p@srwidth{\number\ps@dimcent}%
}%
\def\@p@@sangle#1{%
                \use@psfigtrue
                \@angletrue
                \edef\@p@sangle{#1}%
}%
\def\@p@@ssilent#1{%
                \@verbosefalse
}%
\def\@p@@snoisy#1{%
                \@verbosetrue
}%
\def\@cs@name#1{\csname #1\endcsname}%
\def\@setparms#1=#2,{\@cs@name{@p@@s#1}{#2}}%
\def\ps@init@parms{%
                \@bbllxfalse \@bbllyfalse
                \@bburxfalse \@bburyfalse
                \@heightfalse \@widthfalse
                \@rheightfalse \@rwidthfalse
                \def\@p@sbbllx{}\def\@p@sbblly{}%
                \def\@p@sbburx{}\def\@p@sbbury{}%
                \def\@p@sheight{}\def\@p@swidth{}%
                \def\@p@srheight{}\def\@p@srwidth{}%
                \def\@p@sangle{0}%
                \def\@p@sfile{}%
                \use@psfigfalse
                \@prologfilefalse
                \def\@sc{}%
                \if@noisy
                        \@verbosetrue
                \else
                        \@verbosefalse
                \fi
                \@clipfalse
}%
\def\parse@ps@parms#1{%
                \@psdo\@psfiga:=#1\do
                   {\expandafter\@setparms\@psfiga,}%
\if@prologfile
\fi
}%
\def\bb@missing{%
        \if@verbose
            \typeout{psfig: searching \@p@sfile \space  for bounding box}%
        \fi
        \epsfgetbb{\@p@sfile}%
        \ifepsfbbfound
            \ps@dimcent=\epsfllx bp\edef\@p@sbbllx{\number\ps@dimcent}%
            \ps@dimcent=\epsflly bp\edef\@p@sbblly{\number\ps@dimcent}%
            \ps@dimcent=\epsfurx bp\edef\@p@sbburx{\number\ps@dimcent}%
            \ps@dimcent=\epsfury bp\edef\@p@sbbury{\number\ps@dimcent}%
        \else
            \epsfbbfoundfalse
        \fi
}
\newdimen\p@intvaluex
\newdimen\p@intvaluey
\def\rotate@#1#2{{\dimen0=#1 sp\dimen1=#2 sp
                  \global\p@intvaluex=\cosine\dimen0
                  \dimen3=\sine\dimen1
                  \global\advance\p@intvaluex by -\dimen3
                  \global\p@intvaluey=\sine\dimen0
                  \dimen3=\cosine\dimen1
                  \global\advance\p@intvaluey by \dimen3
                  }}%
\def\compute@bb{%
                \epsfbbfoundfalse
                \if@bbllx\epsfbbfoundtrue\fi
                \if@bblly\epsfbbfoundtrue\fi
                \if@bburx\epsfbbfoundtrue\fi
                \if@bbury\epsfbbfoundtrue\fi
                \ifepsfbbfound\else\bb@missing\fi
                \ifepsfbbfound\else
                \@latexerr{ERROR. cannot locate BoundingBox}\@whattodobb
                        \@p@@sbbllx{100bp}%
                        \@p@@sbblly{100bp}%
                        \@p@@sbburx{200bp}%
                        \@p@@sbbury{200bp}%
                        \no@bbtrue
                        \psdraft
                \fi
                \count203=\@p@sbburx
                \count204=\@p@sbbury
                \advance\count203 by -\@p@sbbllx
                \advance\count204 by -\@p@sbblly
                \edef\ps@bbw{\number\count203}%
                \edef\ps@bbh{\number\count204}%
                 \edef\@bbw{\number\count203}%
                \edef\@bbh{\number\count204}%
               \if@angle
                        \Sine{\@p@sangle}\Cosine{\@p@sangle}%

{\ps@dimcent=\maxdimen\xdef\r@p@sbbllx{\number\ps@dimcent}%

\xdef\r@p@sbblly{\number\ps@dimcent}%

\xdef\r@p@sbburx{-\number\ps@dimcent}%

\xdef\r@p@sbbury{-\number\ps@dimcent}}%
                        \def\minmaxtest{%
                           \ifnum\number\p@intvaluex<\r@p@sbbllx
                              \xdef\r@p@sbbllx{\number\p@intvaluex}\fi
                           \ifnum\number\p@intvaluex>\r@p@sbburx
                              \xdef\r@p@sbburx{\number\p@intvaluex}\fi
                           \ifnum\number\p@intvaluey<\r@p@sbblly
                              \xdef\r@p@sbblly{\number\p@intvaluey}\fi
                           \ifnum\number\p@intvaluey>\r@p@sbbury
                              \xdef\r@p@sbbury{\number\p@intvaluey}\fi
                           }%
                        \rotate@{\@p@sbbllx}{\@p@sbblly}%
                        \minmaxtest
                        \rotate@{\@p@sbbllx}{\@p@sbbury}%
                        \minmaxtest
                        \rotate@{\@p@sbburx}{\@p@sbblly}%
                        \minmaxtest
                        \rotate@{\@p@sbburx}{\@p@sbbury}%
                        \minmaxtest

\edef\@p@sbbllx{\r@p@sbbllx}\edef\@p@sbblly{\r@p@sbblly}%

\edef\@p@sbburx{\r@p@sbburx}\edef\@p@sbbury{\r@p@sbbury}%
                \fi
                \count203=\@p@sbburx
                \count204=\@p@sbbury
                \advance\count203 by -\@p@sbbllx
                \advance\count204 by -\@p@sbblly
                \edef\@bbw{\number\count203}%
                \edef\@bbh{\number\count204}%
}%
\def\in@hundreds#1#2#3{\count240=#2 \count241=#3
                     \count100=\count240        
                     \divide\count100 by \count241
                     \count101=\count100
                     \multiply\count101 by \count241
                     \advance\count240 by -\count101
                     \multiply\count240 by 10
                     \count101=\count240        
                     \divide\count101 by \count241
                     \count102=\count101
                     \multiply\count102 by \count241
                     \advance\count240 by -\count102
                     \multiply\count240 by 10
                     \count102=\count240        
                     \divide\count102 by \count241
                     \count200=#1\count205=0
                     \count201=\count200
                        \multiply\count201 by \count100
                        \advance\count205 by \count201
                     \count201=\count200
                        \divide\count201 by 10
                        \multiply\count201 by \count101
                        \advance\count205 by \count201
                     \count201=\count200
                        \divide\count201 by 100
                        \multiply\count201 by \count102
                        \advance\count205 by \count201
                     \edef\@result{\number\count205}%
}%
\def\compute@wfromh{%
                \in@hundreds{\@p@sheight}{\@bbw}{\@bbh}%
                \edef\@p@swidth{\@result}%
}%
\def\compute@hfromw{%
                \in@hundreds{\@p@swidth}{\@bbh}{\@bbw}%
                \edef\@p@sheight{\@result}%
}%
\def\compute@handw{%
                \if@height
                        \if@width
                        \else
                                \compute@wfromh
                        \fi
                \else
                        \if@width
                                \compute@hfromw
                        \else
                                \edef\@p@sheight{\@bbh}%
                                \edef\@p@swidth{\@bbw}%
                        \fi
                \fi
}%
\def\compute@resv{%
                \if@rheight \else \edef\@p@srheight{\@p@sheight} \fi
                \if@rwidth \else \edef\@p@srwidth{\@p@swidth} \fi
}%
\def\compute@sizes{%
        \if@scalefirst\if@angle
        \if@width
           \in@hundreds{\@p@swidth}{\@bbw}{\ps@bbw}%
           \edef\@p@swidth{\@result}%
        \fi
        \if@height
           \in@hundreds{\@p@sheight}{\@bbh}{\ps@bbh}%
           \edef\@p@sheight{\@result}%
        \fi
        \fi\fi
        \compute@handw
        \compute@resv
					           \EPS@Width=\@bbw
																\divide\EPS@Width by 1000
   												 \EPS@xscale=\@p@swidth \divide \EPS@xscale by \EPS@Width
					           \EPS@Height=\@bbh
																\divide\EPS@Height by 1000
   												 \EPS@yscale=\@p@sheight \divide \EPS@yscale by\EPS@Height
  \ifnum\EPS@xscale>\EPS@yscale\EPS@xscale=\EPS@yscale\fi
}
\def\psfig{\begingroup\@minisanitize\@@@psfig}
\def\epsfig{\begingroup\@minisanitize\@@@epsfig}

\def\@minisanitize{\@makeother\_\@makeother\:\@makeother\.\@makeother\$}

\def\@@@psfig#1{\vbox {%
        \ps@init@parms
        \parse@ps@parms{#1}%
        \ifnum\@psdraft=1
                \typeout{[\@p@sfilefinal]}%
                \if@verbose
                        \typeout{epsfig: using PSFIG macros}%
                \fi
                \psfig@method
        \else
                \epsfig@draft
        \fi
}
\endgroup
}%

\def\@@@epsfig#1{\vbox {%
        \ps@init@parms
        \parse@ps@parms{#1}%
        \ifnum\@psdraft=1
          \if@angle\use@psfigtrue\fi
          {\ifnum\fig@driver=1\global\use@psfigtrue\fi}%
          {\ifnum\fig@driver=3\global\use@psfigtrue\fi}%
          {\ifnum\fig@driver=4\global\use@psfigtrue\fi}%
          {\ifnum\fig@driver=5\global\use@psfigtrue\fi}%
                \ifuse@psfig
                        \if@verbose
                                \typeout{epsfig: using PSFIG macros}%
                        \fi
                        \psfig@method
                \else
                        \if@verbose
                                \typeout{epsfig: using EPSF macros}%
                        \fi
                        \epsf@method
                \fi
        \else
                \epsfig@draft
        \fi
}
\endgroup
}%

\def\epsf@method{%
        \epsfbbfoundfalse
        \if@bbllx\epsfbbfoundtrue\fi
        \if@bblly\epsfbbfoundtrue\fi
        \if@bburx\epsfbbfoundtrue\fi
        \if@bbury\epsfbbfoundtrue\fi
        \ifepsfbbfound\else\epsfgetbb{\@p@sfile}\fi
        \ifepsfbbfound
           \typeout{<\@p@sfilefinal>}%
           \epsfig@gofer
        \else
          \@latexerr{ERROR - Cannot locate BoundingBox}\@whattodobb
          \@p@@sbbllx{100bp}%
          \@p@@sbblly{100bp}%
          \@p@@sbburx{200bp}%
          \@p@@sbbury{200bp}%
                \count203=\@p@sbburx
                \count204=\@p@sbbury
                \advance\count203 by -\@p@sbbllx
                \advance\count204 by -\@p@sbblly
                \edef\@bbw{\number\count203}%
                \edef\@bbh{\number\count204}%
          \compute@sizes
          \epsfig@@draft
       \fi
}%
\def\psfig@method{%
        \compute@bb
        \ifepsfbbfound
          \compute@sizes
          \psfig@start
          \vbox to \@p@srheight true sp{\hbox to \@p@srwidth true
            sp{\hss}\vss\psfig@end}%
        \else
           \epsfig@draft
        \fi
}%
\def\epsfig@draft{\compute@bb\compute@sizes\epsfig@@draft}%
\def\epsfig@@draft{%
\typeout{<(draft only) \@p@sfilefinal>}%
\if@draftbox
        \hbox{\fbox{\vbox to \@p@srheight true sp{%
        \vss\hbox to \@p@srwidth true sp{ \hss
           {\tt\@p@sfilefinal}
                          \hss }\vss
        }}}%
\else
        \vbox to \@p@srheight true sp{%
        \vss\hbox to \@p@srwidth true sp{\hss}\vss}%
\fi
}%
\psfigdriver{dvips}%
\epsfigRestoreAt

\newcommand{\be}{\begin{equation}}
\newcommand{\ee}{\end{equation}}

\newcommand{\bear}{\begin{eqnarray}}
\newcommand{\eear}{\end{eqnarray}}

\newcommand{\mod}{\left|\hspace{-0.4cm}\begin{array}{c}\\ 
\end{array}\right.}

\newcommand{\corrl}{\left[\hspace{-0.4cm}\begin{array}{c}\\ 
\\
\end{array}\right.}
\newcommand{\corrr}{\left]\hspace{-0.4cm}\begin{array}{c}\\ 
\\
\end{array}\right.}

\def\IJMPA #1 #2 #3 {Int.~J.~Mod.~Phys.~{\bf A#1}\ (19#2) #3}
\def\MPLA #1 #2 #3 {Mod.~Phys.~Lett.~{\bf A#1}\ (19#2) #3}
\def\NPB #1 #2 #3 {Nucl.~Phys.~{\bf B#1}\ (19#2) #3}
\def\PLB #1 #2 #3 {Phys.~Lett.~{\bf B#1}\ (19#2) #3}
\def\PR #1 #2 #3 {Phys.~Rep.~{\bf#1}\ (19#2) #3}
\def\PRD #1 #2 #3 {Phys.~Rev.~{\bf D#1}\ (19#2) #3}
\def\PTP #1 #2 #3 {Prog.~Theor.~Phys.~{\bf #1}\ (19#2) #3}
\def\PRL #1 #2 #3 {Phys.~Rev.~Lett.~{\bf#1}\ (19#2) #3}
\def\RMP #1 #2 #3 {Rev.~Mod.~Phys.~{\bf#1}\ (19#2) #3}
\def\ZPC #1 #2 #3 {Z.~Phys.~{\bf C#1}\ (19#2) #3}

\begin{document}

\begin{titlepage}

\title{\bf BOSONIC THERMAL MASSES IN SUPERSYMMETRY}

\author{
{\bf D. Comelli}\thanks{Work supported by Ministerio de 
Educaci\'on y Ciencia (Spain).}\\ Instituto de F\'{\i}sica Corpuscular - 
IFIC/CSIC \\ Dept. de F\'{\i}sica Te\`orica, Universidad de Val\`encia\\
46100 Burjassot, Val\`encia. Spain\\
and\\
{\bf J.R. Espinosa} \thanks{Work supported by the Alexander-von-Humboldt 
Stiftung.} \\ Deutsches Elektronen Synchrotron DESY. \\
Notkestrasse 85.\ \ 22603 Hamburg. Germany}

\date{} 
\maketitle
\vspace{.5cm}
\def\baselinestretch{1.15}
\begin{abstract}
\noindent Effective thermal masses of bosonic particles in a plasma play 
an important role in many different phenomena. We compute them 
in general supersymmetric models at leading order.  The origin 
of different corrections  is explicitly shown for the formulas 
to be applicable when some particles decouple. The correct 
treatment of Boltzmann decoupling in the presence of trilinear 
couplings and mass mixing is also discussed. As a relevant 
example, we present results for the Minimal Supersymmetric 
Standard Model.  
\end{abstract}
\vspace{2cm}
\leftline{June 1996}

\thispagestyle{empty}

\vskip-20.cm
\rightline{{\bf DESY 96-114}}
\rightline{{\bf FTUV/96-37}}
\rightline{{\bf IFIC/96-45}}
\rightline{{\bf IEM--FT--134/96}}
\rightline{{\bf hep-ph/9606438}}
\vskip3in

\end{titlepage}

\def\baselinestretch{1.1}

\section{Introduction}
\vspace{0.5cm}

If supersymmetry is realized in nature it would have manifold 
and very important implications on the history of the early Universe. In 
fact, much effort has been devoted to the study of supersymmetric cosmology 
and truly supersymmetric solutions to old cosmological problems have been 
proposed (while new problems have also arisen. See \cite{susycosm} for 
review and references). However the (weak scale\footnote{We concentrate 
here on temperatures of that order, relevant for example in studies of 
the electroweak phase transition.}) supersymmetric generalization of the 
Standard Model (SM) is not uniquely defined. First of all, the 
introduction of arbitrary soft supersymmetry breaking parameters, to 
prevent the mass degeneracy between ordinary and supersymmetric particles, 
generates a lot of freedom and second of all there are various options 
related to the particle content and the gauge group definition. This limits
the generality of the predictions that can be made although still permits
to confront different classes of models and theoretical assumptions by 
examining their cosmological implications.

At the high temperatures of the early Universe, supersymmetric particles
would be thermally pair created and would populate the plasma. One of the
simplest consequences of this fact is that the effective thermal mass of 
a generic particle immersed in that plasma would be changed due to 
interactions with supersymmetric ambient particles. It is obvious that 
knowledge of these effective thermal masses is fundamental to describe 
the behaviour and properties of the plasma. 
Moreover, it is well known that these quantities play a crucial role in 
many interesting aspects of the evolution of the early Universe. Various 
examples follow.

In the case of gauge vector bosons  (see e.g. \cite{gross,smilga}) the 
effective thermal mass for longitudinal components corresponds 
to the usual Debye mass, i.e. the inverse screening length of electric 
potentials in the plasma. At leading order (one loop in perturbation 
theory) it is $m_D\sim gT$, where $g$  is the corresponding gauge 
coupling constant. Transverse components have  instead zero thermal mass 
at leading order. For abelian gauge bosons this is true also to all 
orders, corresponding to the non screening of magnetic fields, but for 
non-abelian gauge bosons a magnetic mass of order $g^2 T$ is expected to 
appear non-perturbatively. Supersymmetric particles in the plasma will 
have an influence on Debye masses (see e.g. \cite{grundberg}). In this 
paper we will consider only thermal masses at leading order so that 
magnetic masses will be taken to be zero. 

Let us turn now to scalar thermal masses.
As pointed out by Kirzhnits and Linde \cite{linde}, spontaneously broken 
symmetries are generally restored at high temperatures (see also 
\cite{tres}). This can be 
understood in terms of the effective thermal mass of the (Higgs) scalars 
driving the symmetry breaking. Consider as a particularly relevant example 
the electroweak gauge symmetry. Call $\phi$ the Higgs field responsible of
the breaking. The one-loop approximation for the effective potential of
$\phi$, including the effects of finite temperature is of the form
\be
V(\phi,T)=\frac{1}{2}(\kappa 
T^2-m^2)\phi^2-ET(\phi^2)^{3/2}+\frac{1}{4}\lambda(T)\phi^4, \label{V}
\ee
where $E$, $\kappa$ and $\lambda(T)$ are some functions of the masses and 
couplings, easily calculable in a given model. For low temperatures the 
negative $T=0$ mass squared dominates favoring the formation of a 
condensate, while at sufficiently high T, the (leading order) Higgs thermal 
mass $\sqrt{\kappa} T \sim [coupling]\times T$  dominates over the 
negative $T=0$ mass disfavoring a non-zero condensate. Furthermore, it 
is clear that knowledge of the thermal mass for the Higgses allows an 
estimate of the critical temperature  of the transition ($T_c^2\sim 
m^2/\kappa$).

The order of such transitions is also related to the value of thermal masses
in a more indirect way. From (\ref{V}) it is clear that the presence of 
the non-analytic cubic term causes the transition to be first order. In 
fact, the jump in the order parameter is
\be
\frac{\phi(T_c)}{T_c}=\frac{2E}{\lambda(T_c)},
\label{jump}
\ee
[here $T_c$ is defined by the coexistence of two-degenerate vacua in 
(\ref{V})]. The quantity (\ref{jump}) is of the utmost importance for the
viability of electroweak baryogenesis (for review and references see e.g. 
\cite{ewbaryo}). 
Now, the cubic term in (\ref{V}) is a purely finite temperature effect
and comes from the interaction of the Higgs field with the static modes of 
different species of bosons in the plasma (fermions do not contribute
to this term because they do not have static modes). In fact, each 
bosonic degree of freedom, with ($T=0$) field dependent mass $M_i(\phi)$ 
contributes to the potential $V(\phi,T)$ a term
\be
\Delta_i V=-\frac{T}{12\pi}[M_i^2(\phi)]^{3/2}.
\label{cube}
\ee
Beyond the one-loop approximation for the potential, every mass $M_i(\phi)$
in (\ref{cube}) should be substituted by the corresponding effective thermal 
mass, obtained from
\be
M_{ij}^2(\phi)\rightarrow M_{ij}^2(\phi) + \kappa_{ij} T^2,
\label{thermass}
\ee
where the last piece comes from the interaction of the particles
with the surrounding plasma.
Substitution of (\ref{thermass}) in (\ref{cube}) resums an infinite 
series of higher order diagrams, the so-called Daisies. The net effect
of this resummation is to screen the cubic term in (\ref{V}), effectively
reducing the $E$ parameter and thus weakening the strength of the phase 
transition. 
In the Standard Model, where the dominant contribution to the cubic term 
in the potential comes from gauge bosons, the screening of the longitudinal 
modes is very effective while it is zero at leading order for the transverse 
modes. Then, daisy improvement of the effective potential leads to a reduction 
of the strength of the transition \cite{smpht} roughly by a factor 2/3.
In the Minimal Supersymmetric Standard Model (MSSM) if stops are light they
give the dominant contribution to the cubic term in the Higgs potential 
(see \cite{mssmpht} for the effect of Debye screening on the electroweak 
phase transition in the MSSM) and the final strength of the transition will 
be sensitive to the value of stop thermal masses.

Besides the effects explained, (bosonic) thermal mass 
corrections are very important because they represent the starting point 
of a resummation of perturbation theory (see e.g. \cite{tres,t}). Such 
resummation is necessary to 
take care of the infrared problems that plague theories at finite temperature
if they contain massless bosons in the symmetric phase, e.g. 
Yang-Mills theories \cite{lind}. The problem appears when we probe our 
system to low 
scales\footnote{Here $g$ stands for a typical gauge coupling or a Yukawa 
coupling. For power counting quartic scalar couplings are $\lambda\sim 
g^2$.} 0($gT$) compared with the temperature T. At this scale, an infinite 
number of diagrams can give contributions of the same order and to improve 
the usual perturbative series they need to be resummed. The effective 
thermal masses provide then an IR cut-off taming the perturbative 
expansion\footnote{Of course, there remains an infrared problem for 
transverse gauge bosons, associated with physics at the scale $g^2 T$.}.
One example is provided by the cubic term  in the potential discussed 
previously. Its non-analytic behaviour signals its infrared singular origin: 
it comes from (bosonic) zero Matsubara frequency modes. Note that 
fermions do not cause infrared problems because they do not have zero 
Matsubara  modes. In fact, at sufficiently high temperatures (or for 
distances much larger than $1/T$), fermions decouple from the effective 
3D theory at finite T.  For that reason we concentrate here on bosons only.

Other examples where effective thermal masses play a role 
(in supersymmetric contexts) are:
studies on the 3D reduced effective theory in 
the MSSM \cite{mssmred}, analysis of charge and color 
breaking minima \cite{ccb} at finite 
temperature \cite{ccbt}, non-restoration of symmetries at very high 
temperature in general supersymmetric models \cite{nonres}, inverse symmetry
breaking at some range of temperatures \cite{isb}-\cite{cpb}, different
details of the spontaneous mechanism for electroweak baryogenesis 
\cite{spont}, etc.

The aim of this paper is then to compute thermal masses for bosons 
(scalars or gauge vectors) in general softly-broken supersymmetric 
models (section 2). In subsection 2.1 these masses are presented for 
temperatures  much larger than all particle masses. In that case all 
particles in the theory are thermally produced and form part of the 
plasma.
In subsection 2.2 we present the more complicated case in which the 
temperature is lower than the mass of some particles which decouple
from the thermal plasma and then do not contribute to the effective
masses of other particles. 
Section 3 applies these results to the particular case of the MSSM 
(some of the results presented have already appeared in the 
literature \cite{grundberg,mssmpht,pola,polar,polari}).

\section{General Softly-Broken Supersymmetric Model}
\vspace{0.5cm}

Since Bose-Einstein and Fermi-Dirac distributions are different, the 
thermal bath is populated by different amounts of on shell fermions and 
bosons. In such a sense, in a SUSY theory, temperature effects can 
invalidate  various cancellations implied by the symmetry between 
fermions and bosons \cite{salo}. This observation is very relevant in 
particular for 
the computation of effective thermal masses. 

As is well known, only self-energy diagrams which are quadratically 
divergent at $T=0$ contribute to the leading thermal masses. Typical 
diagrams that enter such calculation are depicted in figure~1. Although 
the second diagram is not quadratically divergent it can give a 
contribution in the presence of Boltzmann decoupling and should be kept. 
Note that for our purpose external momentum can be set to zero. 
\begin{figure}[hbt]
\centerline{
\psfig{figure=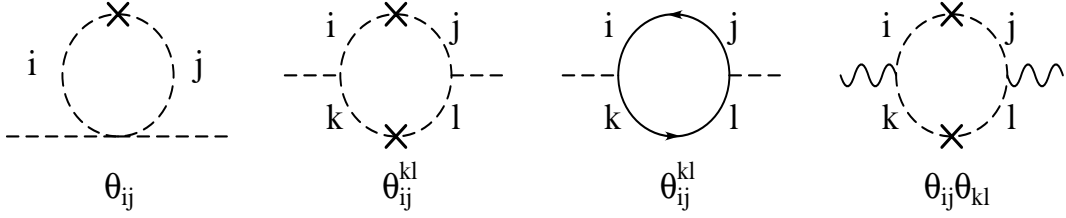,height=4cm,width=8cm,bbllx=5.cm,bblly=17.5cm,bburx=16.cm,bbury=22.5cm}}
\caption{\footnotesize Different types of diagrams contributing 
to thermal masses and responsible for the indicated $\theta$ symbols.} 
\end{figure} 
If fermion-boson cancellations
were still operative at non zero temperature in the supersymmetric case
we would obtain zero thermal masses. However, it can be shown that 
fermionic contributions come with an extra factor $(-1/2)$. More 
explicitly, if a bosonic integral gives
\be
I_b=\kappa(\Lambda^2 + T^2)+...
\ee
where $\Lambda^2$ is the $T=0$ quadratic divergence and  $\kappa T^2$ the
associated finite temperature contribution to the thermal mass, the fermionic
counterpart will be
\be
I_f=-\kappa(\Lambda^2 - \frac{1}{2}T^2)+...
\ee
Then, instead of cancellation of thermal masses there is a reinforcement:
\be
I_b+I_f=\frac{3}{2}\kappa T^2+...
\ee
Explicit examples of this effect can be found in the next sections.

\subsection{Thermal masses in the limit $T\gg M$}
\vspace{0.5cm}

As we will see, in the limit that the temperature is much larger than 
any mass in the theory the contributions to the various self energies 
depend only on the gauge structure of the theory and the dimensionless 
parameters of the superpotential W  which reads 
\be
W=\frac{1}{2}\mu_{ij}\phi_i\phi_j+\frac{1}{3!}W_{ijk}\phi_i\phi_j\phi_k.
\ee
Latin indices $i,j,k,...$ will be used for scalar fields. The 
corresponding fermionic partners carry a tilde: $\tilde{k},\tilde{l},...$. 
Latin indices $a,b,c,...$ are reserved for gauge bosons and the tilded 
version for gauginos. Unless stated otherwise sum over repeated indices 
is always implied.

The full scalar potential is then:
\bear
V_0(\phi)&=& \sum_{i} \mod\frac{\partial W}{\partial \phi_i}\mod^2+
\frac{1}{2}g_a^2 \mod \phi_i^* T^a_{ij} \phi_j\mod^2 \nonumber\\
&+& m_i^2 |\phi_i|^2+ \frac{1}{2}B_{ij}\phi_i\phi_j  + 
\frac{1}{3!}A_{ijk}\phi_i\phi_j\phi_k. 
\eear
The remaining soft breaking terms\footnote{If the model contains matter 
fermions in the adjoint representation of some group, soft masses that 
mix them with the corresponding gauginos can be written. However these 
soft terms are generically absent in supergravity scenarios.}
 are gaugino masses:
\be
\tilde{V}_a M_{\tilde{a}} \tilde{V}_a+h.c.
\ee

Leading order thermal masses for scalars can be obtained simply taking 
derivatives from the one-loop finite T effective potential which reads
\be
V(\phi)=V_0(\phi)+V_1(\phi)+V_T(\phi),
\ee
where $V_1(\phi)$ is the $T=0$ one-loop correction and 
\be
V_T(\phi)=\frac{T^4}{2 \pi^2} \sum_s g_s 
J_s\left(\frac{m_a^2(\phi)}{T^2}\right), 
\ee
with $J_s=J_+(J_-)$  if the $s^{th}$ particle is a boson (fermion) with 
$g_s$ degrees of freedom (defined negative for fermions) and 
\be
J_{\pm}(y^2)= \int^{\infty}_0 dx x^2 \log \left[ 1-(\pm) 
e^{-\sqrt{x^2+y^2}}\right]. 
\ee
   
The behaviour of $J_{\pm}(m^2/T^2)$ is very simple in two different limit 
cases. First of all, the expansion of $J_{\pm}(m^2/T^2)$ for large values 
of $m/T$ give contributions that are exponentially suppressed $\sim 
e^{-m/T}$, while the expansion for small values of $m/T$ gives the
leading contributions $0(T^2)$:
\be
V_T\sim 
\frac{T^2}{24}\sum_{boson}g_b m_b^2+\frac{T^2}{48}\sum_{fermion}g_fm_f^2+...
\label{pott}
\ee
Then, we will simply use a step approximation for the effective potential to 
compute the thermal mass corrections:
\be
V_T=\frac{T^2}{24}\sum_{boson}g_b m_b^2 
\theta_b+\frac{T^2}{48}\sum_{fermion}g_f m_f^2\theta_f 
\label{poth}
\ee 
where the sum is on all mass eigenstates calculated in the theory at zero 
temperature and  $\theta_{b,f}=1$ if $m_{b,f}\ll T$ and 0 if $m_{b,f}\gg 
T$. Of course this is a crude approximation but gives the correct results 
in the two limiting cases of interest.
Now we study the limit in which the temperature is the 
larger mass scale. In such a case $\theta_{b,f}=1$ for all bosons  
and fermions. The analysis is very simplified due to the fact
that we can forget the complications induced by soft breaking 
mass terms and supersymmetric massive parameters present in the 
superpotential.

Setting then all $\theta$'s to 1 in (\ref{poth}) and using the fact that
\be
Str M^2(\phi)\equiv 3TrM_V^2+Tr M_S^2-2TrM_F^2
=\sum_s g_s m_s^2(\phi)= K - 2 g_aD^aTrT^a,
\ee
with $K$ a field independent constant and $D^a=\phi_i^*T^a_{ij}\phi_j$, we 
obtain 
\be
V_T\sim \frac{T^2}{16}\sum_fg_fm_f^2+D-term+K'.
\label{pots}
\ee
The constant term is irrelevant for our purposes and the D-term 
vanish if we further assume $TrY=0$ for any $U(1)$ gauge group present.
 
Then, up to some irrelevant constant, we can write
\be
V_T(\phi_i,T)=\frac{T^2}{16}\sum_fg_fm_f^2
=\frac{T^2}{8}\corrl
\sum_{i,k}\mod 
W_{ik}\mod^2+4\sum_ag_a^2\sum_{i,k}\phi_i^*(T^aT^a)_{ik}\phi_k\corrr.
\label{vt}
\ee
And from this,
\bear
\Pi_{ij}&=&\frac{\partial^2 V_T(\phi,T)}{\partial \phi_i\partial 
\phi^*_j}=\frac{T^2}{8} \corrl\sum_{k,l} 
W_{ikl}W_{jkl}^*+4\sum_ag_a^2(T^aT^a)_{ij}\corrr.
\label{pol}
\eear
Writing $(T^aT^a)_{ij}=C_a(R)\delta_{ij}$ and using a convenient basis 
for the fields $\phi_i$ we get
\be
\label{pisca}
\Pi_{ij}=\delta_{ij}\frac{T^2}{8} 
\corrl\sum_{k,l}\mod W_{ikl}\mod^2+4\sum_ag_a^2C_a(R_i)\corrr,
\ee
which gives the thermal mass corrections for scalars. This diagonal 
correction should be added to the $T=0$ mass matrix. The eigenvalues of 
this thermally corrected matrix are the end point of our calculation.

The leading order thermal masses for longitudinal gauge bosons, $\Pi_V$, 
get contributions from scalar, fermion and gauge boson loops plus their
supersymmetric partners (note that we can describe the chiral 
supermultiplet contributions either as $S+{\tilde S}$ or ${\tilde F}+F$. 
We use both below in the understanding that no $F$ corresponds to any 
${\tilde S}$) 
\be
\Pi_V=\Pi_V^{(S)}+\Pi_V^{({\tilde 
S})}+ \Pi_V^{(F)}+\Pi_V^{({\tilde F})}+
\Pi_V^{(V)}+\Pi_V^{({\tilde V})}
=\frac{3}{2}\Pi_V^{(S)}+3\Pi_V^{(F)}+\frac{3}{2}\Pi_V^{(V)},
\ee
where the last equality follows from supersymmetry as explained above.
Now, the vector contribution is
\bear
U(1):\; \Pi_V^{(V)}=0,&\;\;\;\;\;\;&
SU(N):\; \Pi_V^{(V)}=\frac{N}{3}g_N^2T^2,
\eear
and according to our rule gaugino loops contribute half this result.
The contributions from scalars and (chiral) fermions are
\bear
\Pi_V^{(S)}=\frac{1}{3}\sum_{S}g^2t_2(R_{S})T^2,&&
\Pi_V^{(F)}=\frac{1}{6}\sum_{F}g^2t_2(R_{F})T^2,
\eear
with $Tr(T^aT^b)=t_2(R)\delta^{ab}$. 
Sfermion loops contribute twice as much, while higgsinos give only a half.
The contribution from non-chiral fermions is twice larger than that from
chiral ones.

The final result is then
\bear
\Pi_{U(1)}&=&\frac{1}{2}g_1^2T^2\corrl\sum_SY_S^2+\sum_FY_F^2\corrr
=\frac{1}{2}g_1^2T^2\corrl\sum_AY_A^2\corrr,\\
\Pi_{SU(N)}&=&\frac{1}{2}g_N^2T^2\corrl N+\sum_St_2(R_S)+
\sum_Ft_2(R_F)\corrr=
\frac{1}{2}g_N^2T^2\corrl N+\sum_At_2(R_A)\corrr,
\eear
where the index $A$ runs over chiral supermultiplets.

We see explicitly that all possible self energies depend only on the 
gauge quantum numbers of the  spectrum and on the Yukawa couplings 
$W_{ijk}$ that appear in the superpotential.
In practice, at very high temperature the masses of the underlying $T=0$
are irrelevant. In such a case, the computation of the leading thermal 
corrections is simplified and they can be derived directly from an exactly 
conformal supersymmetric theory.
 
\subsection{Thermal Masses for general T}
\vspace{0.5cm}

The study of the case in which the scale of the temperature is not the 
dominant one is a bit more involved. The mass scales present in the 
theory, aside from possible  non zero background fields, are the soft 
susy breaking terms and the massive  coefficients in the bilinear terms 
of the superpotential. These scales can have very different values and 
there always exist some range of temperatures in which decoupling
and mixing effects have to be taken into account.

The obvious new effect is the Boltzmann 
decoupling of particles of mass $m\gg T$ and then of their contribution 
to thermal masses of other particles. This effect will be taken into 
account by writing every contribution with the corresponding $\theta(T-m)$
that will take care of the decoupling in a step approximation. 

As long as field-background effects can be neglected, i.e. as long as no 
particle is decoupled because of a large background dependent mass, the 
thermal self-energies will only mix particles with the same quantum numbers.
The reason for this is that leading 
thermal masses arise from quadratically divergent diagrams that can 
already be drawn in the $T=0$ unbroken theory.
As an example consider $W_3-B$ mixing. One can certainly draw
diagrams at $T=0$ that mix those particles at one-loop. However when
summing such diagrams over complete $SU(2)$ multiplets these contributions
cancel. At finite T this is reflected in the following non-diagonal 
self-energy:
\bear
\label{w3b}
\Pi_{W_{3L}B_L}&=&\frac{1}{6}g_1g_2T^2\left[
Tr_S(\theta YT_3)+2Tr_F(\theta Y T_3)
\right],
\label{mix}
\eear
which gives zero when all $\theta$'s are 1. In a background that breaks 
$SU(2)\times U(1)$ if some particle acquires a mass larger than T its
contribution to (\ref{mix}) will drop and only then a non-zero contribution 
will result. Here we will always assume that background masses are always
smaller than the temperature (which is usually the case in most 
applications of 
interest) so that we will not encounter this complication. In that case
one can compute thermal masses at zero background (corrections from 
non-zero background effects will be suppressed by powers of $T$). 
The prescription to obtain thermal corrected masses is then to write the 
$T=0$ mass matrices in whatever field background one is interested 
(provided it is smaller than $T$), add 
the thermal corrections and afterwards rotate or diagonalize the mass 
matrix.

Setting then zero background we can in principle compute thermal self 
energies using an interaction basis or a mass eigenstate basis. The 
first option is more convenient and it is simple to rotate to the mass 
basis in particular cases (note that $\theta$'s are naturally defined in the 
mass basis, so that, to decouple some particle the rotation should be 
made). We will express our general results
in terms of some convenient $\theta$ symbols which vary with the origin 
of the contributions as shown in figure~1. The rules to rotate these symbols
from one basis to another are explained below.

The fields in interaction basis, $\phi_i$, can be written as a linear 
combination of the mass eigenstates $\varphi_\alpha$: 
\be
\phi_i=U_i^\alpha\varphi_\alpha,
\ee
where we stress that the unitary matrices $U$ diagonalize the $M^2$ 
mass matrix calculated at zero background and zero temperature.
The symbol $\theta_{ij}$ comes from the contraction of $\phi_i^*-\phi_j$
to close the loop as shown in figure~1. It is defined
by rotating to the mass basis as
\be
\theta_{ij}=\theta_{U_i^\alpha\varphi_\alpha,U_j^\beta\varphi_\beta}=
U_i^{\alpha*}\;\theta_{\alpha\beta}U_j^\beta\;
=U_i^{\alpha*}\;\theta_{\alpha\alpha}\;U_j^\alpha,
\label{rot}
\ee
with
\be
\theta_{\alpha\beta}=\theta_{\alpha\alpha}\delta_{\alpha\beta}=
 \left\{ \begin{array}{c}
1\;\;if\;\;m_{\alpha}\ll T\\
0\;\;if\;\;m_{\alpha}\gg T\\
\end{array}\right.
\ee
The $\theta_{ij}$ symbol defined applies both to fermion or boson 
contractions. 

We define also the 4-index symbol $\theta^{kl}_{ij}$ for the second and 
third diagrams shown in fig.~1. For these objects the rotation from the 
interaction basis to the mass basis is 
\be
\theta^{kl}_{ij}=U^{\beta *}_{i}\;U^\beta_j 
\;\theta^{\alpha\alpha}_{\beta\beta} \;U^{\alpha 
*}_{k}\;U^{\alpha}_l. 
\label{rotbis}
\ee
But now there is a difference for the fermionic and bosonic case.
For fermions we have simply (tildes omitted)
\be
\theta^{\alpha\alpha}_{\beta\beta} =\theta_{\alpha\alpha}\theta_{\beta\beta},
\ee
while for bosons:
\be
\theta_{\beta\beta}^{\alpha\alpha}
=\frac{\theta_{\alpha\alpha}-\theta_{\beta\beta}}{m_\alpha^2-m_\beta^2}.
\label{theb}
\ee
The reason for this is the following: note that the bosonic diagram is not 
quadratically divergent (in particular 
$\theta_{\beta\beta}^{\alpha\alpha}=0$ if 
$\theta_{\alpha\alpha}=\theta_{\beta\beta}=1$). However it contributes to 
the thermal masses
if one of the particles running in the loop, say $\alpha$, decouples. In 
such 
case, the diagram behaves effectively as the first one, with the heavy 
line in the loop collapsed to a point. In other words, in the effective 
theory that results after integrating out the heavy particle there are 
new quartic couplings proportional to $1/m_\alpha^2$. The symbol 
(\ref{theb}) takes this into account.

There is another effect we have to mention before presenting the results.
Suppose that the scalar fields $\phi_i$ and $\phi_j$ have the same 
quantum numbers but opposite abelian charges. The mixing 
$\phi_i-\phi_j^*$ by thermal mass effects is not possible 
in the non-decoupling case analyzed in the previous subsection [see 
 eqs.~(\ref{vt}) and (\ref{pol})] but becomes
possible in the case that thermal contributions from some particles are 
Boltzmann suppressed. We allow for such possibility in our 
general formulas. The corresponding thermal self-energy will be denoted
by $\Pi_{\phi_i,\phi_j^*}$.

Also note that we give our results in terms of thermal polarizations and 
$\theta$'s for complex scalar fields. This assumes that real and imaginary 
components behave in the same way, e.g. they decouple together when some 
mass parameter is made heavy, etc. This is no longer the case in the 
presence of large backgrounds, which we assume not to be the case, or for
singlet fields. In this last case, real and imaginary components can have 
different masses and should be treated separately. Our formalism can 
be trivially generalized to take this possibility into account using
relations like
\be
S=\frac{1}{\sqrt{2}}(S^r+iS_i)\Rightarrow
 \left\{ \begin{array}{c}
\theta_{SS}=\frac{1}{2}\left[\theta_{S^rS^r}+\theta_{S^iS^i}\right]\\
\theta_{SS^*}=\frac{1}{2}\left[\theta_{S^rS^r}-\theta_{S^iS^i}\right]\\
\end{array}\right.
\ee
and so on.

The general results are the following:

{\bf A. Scalars}

A.1 Yukawa contribution from fermion loops:

\be
\Pi_{\phi_i,\phi_j}=
\frac{T^2}{24}
W_{ikl}W_{jrs}^*\theta_{{\tilde r}{\tilde k}}
^{{\tilde s}{\tilde l}}.
\ee

A.2 Yukawa contributions from scalar loops:

\be
\Pi_{\phi_i,\phi_j}=
\frac{T^2}{12}
W_{rik}W_{rjl}^*\theta_{lk},\;\;\;\;\;\;
\Pi_{\phi_i,\phi_j^*}=
\frac{T^2}{24} 
W_{rij}W_{rkl}^*\theta_{kl^*}. 
\ee

A.3 Trilinear contributions:
 
Note that these terms are proportional to $\theta_{kl}^{ij}$ and thus 
give zero in the limit $T\gg M$.
\bear
\Pi_{\phi_i,\phi_j}&=&
\frac{T^2}{24}\left\{
A_{ikl}A^*_{jrs}\theta_{ls}^{kr}\right.
+2W_{ikl}\mu^*_{lm}W^*_{jrs}\mu_{rn}(\theta_{mn}^{sk}+
\theta_{ms^*}^{nk^*})
+W_{mkl}^*\mu_{mi}W_{rsn}\mu_{rj}^*\theta_{nl}^{sk}\\
&+&\left.\left[2 
W_{ikl}\mu^*_{km}W_{nrs}\mu^*_{jr}\theta_{l^*s}^{mn}
+ 2 A_{ikl}W^*_{jrs}\mu_{mr}\theta_{m^*l}^{sk}
+ A_{ikl}W_{rsm}\mu^*_{jr}\theta_{m^*l}^{s^*k}
+(h.c., i \leftrightarrow j)\right]\right\}\nonumber,
\eear
\bear
\Pi_{\phi_i,\phi_j^*}&=&
\frac{T^2}{24}\left\{
A_{ikl}A_{jrs}\theta_{l^*s}^{k^*r}
+2W_{ikl}\mu^*_{km}W_{jrs}\mu^*_{rn}(\theta_{ms}^{ln}+\theta_{mn^*}^{ls^*})
+W_{rkl}^*\mu_{ir}W_{mns}^*\mu_{jm}\theta_{s^*l}^{n^*k}\right.\\
&+&\left.\left[2 A_{ikl}W_{jmn}\mu^*_{mr}\theta_{lr}^{kn^*}
+ A_{ikl}W_{mnr}^*\mu_{jm}\theta_{lr}^{kn}
+2W_{rkl}^*\mu_{ir}W_{jmn}\mu^*_{ms}\theta_{s^*l}^{nk}+
( i \leftrightarrow j)\right]\right\}.\nonumber
\eear

A.4 Gauge contributions from fermion loops:

We change momentarily our 
notation 
from $\phi_i$ to $N_\alpha$ where $N$ refers to a given rep. of the group 
and $\alpha$ to the group index (not to confuse with mass eigenstate 
indices). Of course no change is needed for $U(1)$'s.

\be
\Pi_{N_\alpha,P_\beta}=
\frac{T^2}{6}g_a g_b T^a_{\alpha\gamma} T^b_{\delta\beta}
\theta_{{\tilde b}{\tilde a}}
^{{\tilde \gamma}{\tilde \delta}},\;\;\;\;\;
\Pi_{N_\alpha,P_\beta^*}=
\frac{T^2}{6}g_a g_b T^a_{\alpha \gamma} T^b_{\beta \delta}
\theta_{{\tilde a}{\tilde \delta}}
^{{\tilde b}{\tilde \gamma}}.
\ee

The second contribution can be non-zero only if the model contains matter
fermions in the adjoint representation as discussed in footnote 4.

A.5 Gauge contribution from scalar loops:

The general result is:
\be
\Pi_{N_\alpha,P_\beta}=
\frac{T^2}{12}g^2\left[\delta_{NP}T^a_{\beta\alpha}Tr_S(T^a_{\gamma\delta}
\theta_{\gamma\delta})
+T^a_{\gamma\alpha}T^a_{\beta\delta}\theta_{\gamma\delta}
\right].
\ee
\be
\Pi_{N_\alpha,P_\beta^*}=
\frac{T^2}{12}g^2 \left[ T^a_{\gamma\alpha}T^a_{\delta\beta}
\theta_{\delta^* \gamma} \right].
\ee

For $SU(N)$ with all non-singlet fields in the fundamental rep.

\be
\Pi_{N_\alpha,P_\beta}=
\frac{T^2}{24}g^2\left[2\delta_{NP}T^a_{\beta\alpha}Tr_S(T^a\theta)
+\delta_{\alpha\beta}\theta_{N_\gamma P_\gamma}
-\frac{1}{N}\theta_{\alpha\beta}
\right].
\ee
\be
\Pi_{N_\alpha,P_\beta^*}=
\frac{T^2}{24}g^2\left[\theta_{P_\alpha^*N_\beta}-\frac{1}{N}
\theta_{P_\beta^*N_\alpha}\right].
\ee

For $U(1)_Y$:

\be
\Pi_{\phi_i,\phi_j}=\frac{T^2}{12}g_1^2\left[\delta_{ij}
Y_iTr_S(Y\theta)+Y_iY_j\theta_{ij}\right];\;\;\;
\Pi_{\phi_i,\phi_j^*}=\frac{T^2}{12}g_1^2Y_iY_j\theta_{ij^*}.
\ee

A.6 Gauge contribution from gauge boson loops:

The general result is:
\be
\Pi_{P_\alpha,P_\beta}=
\frac{T^2}{4}g_Ag_BT^a_{\beta\gamma}T^b_{\gamma\alpha}\theta_{ab},
\ee

For $SU(N)$, when $\theta_{ab}=\delta_{ab}\theta_{aa}$:

\be
\Pi_{P_\alpha,P_\beta}=
\frac{T^2}{4}g_N^2C_N(R_P)\delta_{\alpha\beta}\theta_G.
\ee

For $U(1)$:

\be
\Pi_{\phi_i,\phi_j}=\frac{T^2}{4}g_1^2Y_i^2\delta_{ij}\theta_B.
\ee

{\bf B. Gauge Bosons}

As already mentioned, only longitudinal gauge bosons get a non-zero 
thermal mass at leading order. The following thermal polarizations should 
then be understood as polarizations for the temporal components $\Pi_{00}$
of the gauge fields $V^a_0,V^b_0$. 

B.1 Scalar contribution:
\be
\Pi_{ab}=\frac{T^2}{12}g_Ag_B\left[
 \{T^a,T^b\}_{\beta\gamma}\theta_{P_{\beta}P_{\gamma}}+
\left[T^a_{\beta\alpha}T^b_{\delta\gamma}\left(\theta_{M_\beta N_\gamma}
\theta_{N_\delta M_\alpha}-\theta_{M_\beta N_\delta^*}
\theta_{N_\gamma M_\alpha^*}\right)+h.c.\right] 
\right].
 \ee
where in principle two different groups, with coupling constants $g_A$, 
$g_B$ are considered.

For $SU(N)$ and fields in the fundamental rep. $M,P,$ etc:

The general result is:
\bear
\Pi_N&\equiv &\frac{1}{N^2-1}\sum_a\Pi_{aa}=
\frac{T^2}{12}g_N^2\left\{
\sum_M \theta_M+\frac{1}{N^2-1}
\left[\theta_{M_\alpha P_\alpha}\theta_{P_\beta M_\beta}
-\theta_{M_\beta P_\alpha^*}\theta_{M_\alpha^* 
P_\beta}\right.\right.\nonumber\\
&-&\left.\left.\frac{1}{N}\left(\theta_{M_\alpha P_\beta}\theta_{P_\beta 
M_\alpha} -\theta_{M_\beta^* P_\alpha}\theta_{M_\alpha P_\beta^*}
\right)\right]
\right\},
\eear
where $\theta_M\equiv (1/N)\sum_\alpha\theta_{M_\alpha M_\alpha}$.

For U(1):
\be
\Pi_{B_L}=\frac{T^2}{6}g_1^2
\sum_{ij}Y_iY_j\left[\theta_{ij}+\theta_{ij}^2
-\theta_{ij^*}^2\right].
\ee

B.2 Contributions from matter fermion loops:
\be
\Pi_{ab}=\frac{T^2}{6}g_Ag_B 
T^a_{\alpha\beta}T^b_{\gamma\delta}\theta_{{\tilde 
P}_{\beta}{\tilde N}_{\gamma}}\theta_{{\tilde 
P}_{\alpha}{\tilde N}_{\delta}}.
\ee

B.3 Contributions from gaugino loops:
\be
\Pi_{ab}=\frac{T^2}{6}g_Ag_Bf^A_{dea}f^B_{ghb}
\theta_{{\tilde e}{\tilde g}}^{{\tilde d}{\tilde h}},
\ee
where $f^A_{abc}$ are the structure constants of the group A 
($[T^a,T^b]=if_{abc}T^c$).

\section{Minimal Supersymmetric Standard Model}

In this section we apply our general results to a particularly relevant 
example, the MSSM.
It is the simplest supersymmetric extension of the SM and is described
by the superpotential
\be
W=\mu H_1 \cdot H_2 + h_{U_i} Q_i \cdot H_2 U_i+h_{D_i} 
H_1 \cdot Q_iD_i+h_{E_i} H_1 \cdot L_iE_i 
\ee
embedded into the $SU(3)\times SU(2)\times U(1)$ gauge group.
We will retain full freedom in the soft susy breaking terms in order to be as
general as possible. However we will assume negligible intergenerational 
mixing. In such case the only fields that mix at zero background are
$H_1$ and $H_2$.

The scalar-fermionic soft lagrangian reads:
\bear
L_{Soft}&=& A_{E_i}H_1\cdot \tilde{L}_i \tilde{E}_i+ A_{D_i} 
H_1\cdot\tilde{Q}_i \tilde{D}_i+
 A_{U_i} \tilde{Q}_i \cdot H_2\tilde{U}_i\nonumber\\
&-&m_3^2H_1\cdot H_2
+[\sum_{\tilde{g}}\tilde{g}M\tilde{g}+h.c] +\sum_{i}m_{i}^2|\phi_i|^2.
\eear

\subsection{Limit of very large T}

The bosonic self energies in the case in which 
the temperature is the larger mass scale, i.e.  $T\gg\mu,\;
A_{\phi},\;m_{\phi},\;m_3,\;M$ are obtained from subsection~2.1 directly
as

\bear
\Pi_{{\tilde U}_{L_i}}&=&\Pi_{{\tilde D}_{L_i}}=\frac{2}{3}g_3^2T^2+
\frac{3}{8}g_2^2T^2+\frac{1}{72}g_1^2T^2+\frac{1}{4}(h_{U_i}^2+h_{D_i}^2)T^2,
\nonumber\\
\Pi_{{\tilde U}_{R_i}}&=&\frac{2}{3}g_3^2T^2
+\frac{2}{9}g_1^2T^2+\frac{1}{2}h_{U_i}^2T^2,\nonumber\\
\Pi_{{\tilde D}_{R_i}}&=&\frac{2}{3}g_3^2T^2
+\frac{1}{18}g_1^2T^2+\frac{1}{2}h_{D_i}^2T^2,\nonumber\\
\Pi_{{\tilde e}_{L_i}}&=&\Pi_{{\tilde \nu}_{L_i}}=\frac{3}{8}g_2^2T^2
+\frac{1}{8}g_1^2T^2,\nonumber\\
\Pi_{{\tilde e}_{R_i}}&=&\frac{1}{2}g_1^2T^2,\nonumber\\
\Pi_{H_1^0}&=&\Pi_{H_1^\pm}=\frac{3}{8}g_2^2T^2
+\frac{1}{8}g_1^2T^2+\frac{3}{4}h_b^2T^2,\nonumber\\
\Pi_{H_2^0}&=&\Pi_{H_2^\pm}=\frac{3}{8}g_2^2T^2
+\frac{1}{8}g_1^2T^2+\frac{3}{4}h_t^2T^2,\nonumber\\
\Pi_{g_L}&=&\frac{9}{2}g_3^2T^2,\nonumber\\
\Pi_{W_L}&=&\frac{9}{2}g_2^2T^2,\nonumber\\
\Pi_{g_L}&=&\frac{11}{2}g_1^2T^2,\nonumber\\
\eear

For $H_1$ and $H_2$ we only keep third generation Yukawa couplings. 
Also, note that particles in the same gauge multiplet receive the same 
thermal mass correction. 

\subsection{Explicit formulas in the general case}

If there is not a defined hierarchy between the
scales $T,\;\mu,\;
A_{\phi},\;m_{i},\;m_3,\;M$, we must apply the formulas of 
subsection 2.2 that obviously have as asymptotic limit , for high T, the
equations presented in the previous subsection.
 
In the formulas that follow we write most of the self energies in the 
interaction basis. Also, we use $\theta_i=\theta_{ii}$, 
$\theta_i^j=\theta_{ii}^{jj}$, etc, to simplify the notation.
For all fields besides $H_1$ and $H_2$ 
the $\theta_{ij}$ functions in the gauge basis are diagonal and coincide with 
the definition in the mass eigenstate basis (at zero background).
The treatment of $H_{1,2}$ is as follows:
as is well known, there are three mass parameters in the tree-level Higgs
potential of the MSSM:
\be
V=m_1^2|H_1|^2+m_2^2|H_2|^2+m_3^2(H_1\cdot H_2 + h.c.) + quartic\;\; terms.
\ee
Two of these mass parameters ($m_1,m_2$) can be traded by the $T=0$ vacuum 
expectation values $v_1$ and $v_2$ [with  $v_1^2+v_2^2=v^2=(174\ GeV^2)$ 
and $\tan\beta=v_2/v_1$] leaving only one free mass parameter, 
conventionally 
taken to be the mass of the pseudoscalar Higgs, $m_A$.  Then we have two 
scales in the Higgs sector, $v$ (or $M_Z$) and $m_A$, and  the only 
non-trivial case  at finite temperature corresponds to $M_Z\ll T\ll m_A$. 
When $m_A\gg M_Z$ one (linear combination) of the two Higgs doublets is 
heavy  ($\sim m_A$) and one light ($\sim M_Z$).

In order to obtain the mass eigenstates at zero background, 
we can work with the full doublets. Diagonalization of the 
$2\times 2$ matrix defines the mixing angle $\beta_0$.  
In the only non-trivial case with $m_A\gg M_Z$, it's straightforward
to see that $\beta_0\rightarrow \beta$ so that, in this limit 
we can define the doublets $H$ (light) and $\Phi$ (heavy) by the rotation 
\bear
\label{rotH}
{\overline {H}}_1 &=&H\cos\beta  - \Phi\sin\beta ,\nonumber\\
H_2 &=&H\sin\beta  + \Phi\cos\beta ,
\eear
where ${\overline { H}}_1=(-H_1^+, H_1^{0*})^T$. 
The doublets $H$ and $\Phi$ are the mass eigenstates so 
that (\ref{rotH}) is our equation $\phi_i=U^\alpha_i\varphi_\alpha$ in 
this case. From (\ref{rotH}) and using rules (\ref{rot}) and (\ref{rotbis})
we can express all $\theta$ symbols for Higgs bosons in terms of
$\theta_{H}$ and $\theta_\Phi$, or equivalently $\theta(T-M_Z)$ and
$\theta(T-m_A)$.

From (\ref{rotH}) it follows
\bear
\theta_{H_1^0H_1^0}&=&\theta_{H^0H^0}\cos^2\beta
+\theta_{\Phi^0\Phi^0}\sin^2\beta,\nonumber\\
\theta_{H_1^{0*}H_2^0}&=&(\theta_{H^0H^0}
-\theta_{\Phi^0\Phi^0})\cos\beta\sin\beta,\nonumber\\
\theta_{H_1^{-*}H_2^+}&=&-(\theta_{H^+H^+}
-\theta_{\Phi^+\Phi^+})\cos\beta\sin\beta,
\eear
and so on.
The rest of $\theta$ symbols are trivial to handle. For squarks remember
that gauge invariance requires equal soft mass $m_{\tilde Q}$ for 
${\tilde U}_L$ and ${\tilde D}_L$ so that $\theta_{{\tilde U}_L}=
\theta_{{\tilde U}_L}\equiv\theta_{{\tilde Q}}$.

Also note that although $\theta$'s for gauge bosons will always take the 
value 1 (because they are massless at zero background) write them 
explicitly. 

We also use 
\bear
6Tr_S(\theta Y)&=&-3\sum_j(\theta_{{\tilde \nu}_{L_j}}+\theta_{{\tilde 
e}_{L_j}} -2\theta_{{\tilde e}_{R_j}})+
3(\theta_{H_2^\pm}+\theta_{H_2^0})\nonumber\\
&-&
3(\theta_{H_1^\pm}+\theta_{H_1^0})+
N_c\sum_j(\theta_{{\tilde U}_{L_j}}+\theta_{{\tilde 
D}_{L_j}}-4 \theta_{{\tilde U}_{R_j}}+2\theta_{{\tilde D}_{R_j}}),
\eear
\bear
2Tr_S(\theta T_3)&=&\sum_{j}(\theta_{{\tilde \nu}_{L_j}}- \theta_{{\tilde 
e}_{L_j}})+N_c\sum_{j}(\theta_{{\tilde U}_{L_{j}}}- 
\theta_{{\tilde D}_{L_{j}}})
+\theta_{H_1^0} -\theta_{H_1^\pm}+\theta_{H_2^\pm}
-\theta_{H_2^0}.
\eear
Note however, that this last trace would be non-zero only in a $SU(2)_L$
breaking background.

In the formulas that follow the reader can easily check sector by sector
that fermionic contributions are always half of the corresponding bosonic 
ones. 

{\bf A. SQUARKS}

Thermal self energies are diagonal in color space unless a color-breaking
background that decouples some contribution is present. As we assume this 
is not the case the color index structure is trivial and is suppressed.

\bear
\Pi_{{\tilde U}_{L_i}}&=&\frac{1}{6}g_3^2T^2\frac{N_c^2-1}{4N_c}\left[
3\theta_g+\theta_{{\tilde U}_{L_i}}+2\theta_{{\tilde g}}\theta_{U_{L_i}}
\right] \nonumber\\
&+&\frac{1}{48}g_2^2T^2\left[6\theta_{W^\pm}+3\theta_{W_3}
+\theta_{{\tilde U}_{L_i}}+2\theta_{{\tilde D}_{L_i}}
+2 Tr_S(\theta T_3)
+ 2\theta_{{U}_{L_i}}\theta_{{\widetilde 
W}_3} + 4 \theta_{{D}_{L_i}}\theta_{{\widetilde W}^\pm} 
\right]\nonumber\\
&+&\frac{1}{432}g_1^2T^2\left[3\theta_B+
\theta_{{\tilde U}_{L_i}}
+6Tr_S(\theta Y)
+2\theta_{{\tilde B}}\theta_{{U}_{L_i}}\right]
+\Delta_{{\tilde U}_{R_i}}^{0}
+\Delta_{{\tilde D}_{R_i}}^{\pm}\\
&+&\frac{1}{12}h_{U_i}^2T^2\left[\theta_{H_2^0}+
\theta_{{\tilde U}_{R_i}}+
\theta_{{\tilde H}_2^0}\theta_{U_{R_i}}
\right]+\frac{1}{12}h_{D_i}^2T^2\left[\theta_{H_1^\pm}+
\theta_{{\tilde D}_{R_i}}+
\theta_{{\tilde H}_1^\pm}\theta_{D_{R_i}}
\right],\nonumber
\eear
\bear
\Pi_{{\tilde D}_{L_i}}&=&\frac{1}{6}g_3^2T^2\frac{N_c^2-1}{4N_c}\left[
3\theta_g+\theta_{{\tilde D}_{L_i}}+2\theta_{{\tilde g}}\theta_{D_{L_i}}
\right] \nonumber\\
&+&\frac{1}{48}g_2^2T^2\left[6\theta_{W^\pm}+3\theta_{W_3}
+\theta_{{\tilde D}_{L_i}}+2\theta_{{\tilde U}_{L_i}}
-2Tr_S(\theta T_3)
+ 2\theta_{{D}_{L_i}}\theta_{{\widetilde 
W}_3} + 4 \theta_{{U}_{L_i}}\theta_{{\widetilde W}^\pm} 
\right]\nonumber\\
&+&\frac{1}{432}g_1^2T^2\left[3\theta_B+
\theta_{{\tilde D}_{L_i}}
+6Tr_S(\theta Y)
+2\theta_{{\tilde B}}\theta_{{D}_{L_i}}\right]+\Delta_{{\tilde D}_{R_i}}^{0}
+\Delta_{{\tilde U}_{R_i}}^{\pm}\\
&+&\frac{1}{12}h_{U_i}^2T^2\left[\theta_{H_2^\pm}+
\theta_{{\tilde U}_{R_i}}+
\theta_{{\tilde H}_2^\pm}\theta_{U_{R_i}}
\right]+\frac{1}{12}h_{D_i}^2T^2\left[\theta_{H_1^0}+
\theta_{{\tilde D}_{R_i}}+
\theta_{{\tilde H}_1^0}\theta_{D_{R_i}}
\right],\nonumber
\eear
\bear
\Pi_{{\tilde U}_{R_i}}&=&\frac{1}{6}g_3^2T^2\frac{N_c^2-1}{4N_c}\left[
3\theta_g+\theta_{{\tilde U}_{R_i}}+2\theta_{{\tilde g}}\theta_{U_{R_i}}
\right] \nonumber\\
&+&\frac{1}{108}g_1^2T^2\left[12\theta_B+
4\theta_{{\tilde U}_{R_i}}
-6Tr_S(\theta Y)
+8\theta_{{\tilde B}}\theta_{{U}_{R_i}}\right]+\Delta_{{\tilde U}_{L_i}}^{0}
+\Delta_{{\tilde D}_{L_i}}^{\pm}\\
&+&\frac{1}{12}h_{U_i}^2T^2\left[\theta_{H_2^0}+
\theta_{H_2^\pm}+
\theta_{{\tilde U}_{L_i}}+
\theta_{{\tilde D}_{L_i}}+
\theta_{{\tilde H}_2^0}\theta_{U_{L_i}}+
\theta_{{\tilde H}_2^\pm}\theta_{D_{L_i}}\right],\nonumber
\eear
\bear
\Pi_{{\tilde D}_{R_i}}&=&\frac{1}{6}g_3^2T^2\frac{N_c^2-1}{4N_c}\left[
3\theta_g+\theta_{{\tilde D}_{R_i}}+2\theta_{{\tilde g}}\theta_{D_{R_i}}
\right] \nonumber\\
&+&\frac{1}{216}g_1^2T^2\left[6\theta_B+
2\theta_{{\tilde D}_{R_i}}
+6Tr_S(\theta Y)
+4\theta_{{\tilde B}}\theta_{{D}_{R_i}}\right]+\Delta_{{\tilde D}_{L_i}}^{0}
+\Delta_{{\tilde U}_{L_i}}^{\pm}\\
&+&\frac{1}{12}h_{D_i}^2T^2\left[\theta_{H_1^0}+
\theta_{H_1^\pm}+
\theta_{{\tilde U}_{L_i}}+
\theta_{{\tilde D}_{L_i}}+
\theta_{{\tilde H}_1^0}\theta_{D_{L_i}}+
\theta_{{\tilde H}_1^\pm}\theta_{U_{L_i}}\right],\nonumber
\eear
where
\be
\Delta_{\tilde{U}_P}^{c}=\frac{T^2}{12}h_U^2\left[
| A_{U_P}|^2\theta^{\tilde{U}_P}_{H_2^c}+|\mu|^2 
\theta^{\tilde{U}_P}_{H_1^c}+
( A_{U_P} \mu+A_{U_P}^*\mu^*)\theta^{\tilde{U}_P}_{H_1^cH_2^{c*}}
\right],\nonumber
\ee
\be
\Delta_{\tilde{D}_P}^{c}=\frac{T^2}{12}h_D^2\left[
| A_{D_P}|^2\theta^{\tilde{D}_P}_{H_1^c}+
|\mu|^2 \theta^{\tilde{D}_P}_{H_2^c}-
( A_{D_P}\mu+A_{D_P}^*\mu^*)\theta^{\tilde{D}_P}_{H_1^cH_2^{c*}}
\right].
\ee
Rotating from $H_1, H_2$ to $H, \Phi$ as explained, the previous $\Delta$'s
can be written as
\be
\Delta_{P}^{c}=\frac{T^2}{12}h_P^2\left[
|{\tilde A}_P^{+}|^2\frac{\theta_{P}-\theta_{H^c}}{m_P^2-m_Z^2}+
|{\tilde A}_P^{-}|^2\frac{\theta_{P}-\theta_{\Phi^c}}{m_P^2-m_A^2}
\right].
\ee
Here $P={\tilde U}_{L_i},{\tilde D}_{L_i},{\tilde U}_{R_i},{\tilde 
D}_{R_i}$; $c=0,\pm$ and \bear
{\tilde A}_{U_i}^{+}=A_{U_i}\sin\beta+\mu^*\cos\beta,&&
{\tilde A}_{U_i}^{-}=A_{U_i}\cos\beta-\mu^*\sin\beta,\nonumber\\
{\tilde A}_{D_i}^{+}=A_{D_i}\cos\beta+\mu^*\sin\beta,&&
{\tilde A}_{D_i}^{-}=A_{D_i}\sin\beta-\mu^*\cos\beta.\nonumber
\eear

{\bf B. SLEPTONS}

\bear 
\Pi_{{\tilde
\nu}_{L_i}}&=&\frac{1}{48}g_2^2T^2\left[6\theta_{W^\pm}+3\theta_{W_3}
+\theta_{{\tilde \nu}_{L_i}}+ 2\theta_{{\tilde e}_{L_i}}
+ 2Tr_S(\theta T_3)
+ 2\theta_{{\nu}_{L_i}}\theta_{{\widetilde 
W}_3} + 4
\theta_{{e}_{L_i}}\theta_{{\widetilde W}^\pm} \right]\nonumber\\
&+&\frac{1}{144}g_1^2T^2\left[9\theta_B+ 3\theta_{{\tilde \nu}_{L_i}}
-6Tr_S(\theta Y) +6\theta_{{\tilde
B}}\theta_{{\nu}_{L_i}}\right]+\Delta_{{\tilde E}_{R_i}}^{\pm}\\
&+&
\frac{1}{12}h_{E_i}^2T^2\left[\theta_{H_1^\pm}+ \theta_{{\tilde E}_{R_i}}+
\theta_{{\tilde H}_1^\pm}\theta_{E_{R_i}} \right],\nonumber 
\eear

\bear \Pi_{{\tilde
e}_{L_i}}&=&\frac{1}{48}g_2^2T^2\left[6\theta_{W^\pm}+3\theta_{W_3}
+\theta_{{\tilde e}_{L_i}}+ 2\theta_{{\tilde \nu}_{L_i}}
-2Tr_S(\theta T_3)
+ 2\theta_{{e}_{L_i}}\theta_{{\widetilde 
W}_3} + 4
\theta_{{\nu}_{L_i}}\theta_{{\widetilde W}^\pm} \right]
\nonumber\\
&+&\frac{1}{144}g_1^2T^2\left[9\theta_B+ 3\theta_{{\tilde e}_{L_i}}
-6Tr_S(\theta Y) +6\theta_{{\tilde
B}}\theta_{{e}_{L_i}}\right]+\Delta_{{\tilde E}_{R_i}}^{0}
\\ &+&
\frac{1}{12}h_{E_i}^2T^2\left[\theta_{H_1^0}+ \theta_{{\tilde E}_{R_i}}+
\theta_{{\tilde H}_1^0}\theta_{E_{R_i}} \right],\nonumber \eear
\bear
\Pi_{{\tilde E}_{R_i}}&=&\frac{1}{72}g_1^2T^2\left[18\theta_B+
6\theta_{{\tilde E}_{R_i}} +6Tr_S(\theta Y) +12\theta_{{\tilde
B}}\theta_{{E}_{R_i}}\right]+\Delta_{{\tilde E}_{L_i}}^{0}
+\Delta_{{\tilde \nu}_{L_i}}^{\pm}\\
&+&\frac{1}{12}h_{E_i}^2T^2\left[\theta_{H_1^0}+ \theta_{H_1^\pm}+
\theta_{{\tilde \nu}_{L_i}}+ \theta_{{\tilde e}_{L_i}}+ \theta_{{\tilde
H}_1^0}\theta_{e_{L_i}}+ \theta_{{\tilde
H}_1^\pm}\theta_{\nu_{L_i}}\right],\nonumber 
\eear 
where the $\Delta$'s follow the same notation used for squarks and now 
\bear
{\tilde A}_{E_i}^{+}=A_{E_i}\cos\beta+\mu^*\sin\beta,&&
{\tilde A}_{E_i}^{-}=A_{E_i}\sin\beta-\mu^*\cos\beta.
\eear

{\bf C. HIGGS BOSONS}

\bear
\Pi_{H_1^0}&=&\frac{1}{48}g_2^2T^2\left[6\theta_{W^\pm}+3\theta_{W_3}
+\theta_{H_1^0}+2\theta_{H_1^\pm}
+ 2Tr_S(\theta T_3)
+ 2\theta_{{\tilde H}_1^0}\theta_{{\widetilde 
W}_3} + 4 \theta_{{\tilde H}_1^\pm}\theta_{{\widetilde W}^\pm} 
\right]\nonumber\\
&+&\frac{1}{144}g_1^2T^2\left[9\theta_B+3\theta_{H_1^0}
-6Tr_S(\theta Y)
+6\theta_{{\tilde B}}\theta_{{\tilde H}_1^0}\right]+\Delta_1\\
&+&\frac{1}{12}T^2\sum_i\left[ N_c h_{D_i}^2\left(\theta_{{\tilde D}_{L_i}}+
\theta_{{\tilde D}_{R_i}}+
\theta_{D_{L_i}}\theta_{D_{R_i}}
\right)+h_{E_i}^2\left(\theta_{{\tilde e}_{L_i}}+
\theta_{{\tilde E}_{R_i}}+
\theta_{e_{L_i}}\theta_{E_{R_i}}
\right)\right],\nonumber
\eear
\bear
\Pi_{H_1^\pm}&=&\frac{1}{48}g_2^2T^2\left[6\theta_{W^\pm}+3\theta_{W_3}
+\theta_{H_1^\pm}+2\theta_{H_1^0}
-2Tr_S(\theta T_3)
+ 2\theta_{{\tilde H}_1^\pm}\theta_{{\widetilde 
W}_3} + 4 \theta_{{\tilde H}_1^0}\theta_{{\widetilde W}^\pm} 
\right]\nonumber\\
&+&\frac{1}{144}g_1^2T^2\left[9\theta_B+3\theta_{H_1^\pm}
-6Tr_S(\theta Y)
+6\theta_{{\tilde B}}\theta_{{\tilde H}_1^\pm}\right]+\Delta_1\\
&+&\frac{1}{12}T^2\sum_i\left[ N_c h_{D_i}^2\left(\theta_{{\tilde U}_{L_i}}+
\theta_{{\tilde D}_{R_i}}+
\theta_{U_{L_i}}\theta_{D_{R_i}}
\right)+h_{E_i}^2\left(\theta_{{\tilde \nu}_{L_i}}+
\theta_{{\tilde E}_{R_i}}+
\theta_{\nu_{L_i}}\theta_{E_{R_i}}
\right)\right],\nonumber
\eear
\bear
\Pi_{H_2^\pm}&=&\frac{1}{48}g_2^2T^2\left[6\theta_{W^\pm}+3\theta_{W_3}
+\theta_{H_2^\pm}+2\theta_{H_2^0}
+2Tr_S(\theta T_3)
+2\theta_{{\tilde H}_2^\pm}\theta_{{\widetilde 
W}_3} + 4 \theta_{{\tilde H}_2^0}\theta_{{\widetilde W}^\pm} 
\right]\nonumber\\
&+&\frac{1}{144}g_1^2T^2\left[9\theta_B+3\theta_{H_2^\pm}
+6Tr_S(\theta Y)
+6\theta_{{\tilde B}}\theta_{{\tilde H}_2^\pm}\right]+\Delta_2\\
&+&\frac{1}{12}T^2\sum_i N_c h_{U_i}^2\left(\theta_{{\tilde D}_{L_i}}+
\theta_{{\tilde U}_{R_i}}+
\theta_{D_{L_i}}\theta_{U_{R_i}}
\right),\nonumber
\eear
\bear
\Pi_{H_2^0}&=&\frac{1}{48}g_2^2T^2\left[6\theta_{W^\pm}+3\theta_{W_3}
+\theta_{H_2^0}+2\theta_{H_2^\pm}
-2Tr_S(\theta T_3)
+2\theta_{{\tilde H}_2^0}\theta_{{\widetilde 
W}_3} + 4 \theta_{{\tilde H}_2^\pm}\theta_{{\widetilde W}^\pm} 
\right]\\
&+&\frac{1}{144}g_1^2T^2\left[9\theta_B+3\theta_{H_2^0}
+6Tr_S(\theta Y)
+6\theta_{{\tilde B}}\theta_{{\tilde H}_2^0}\right]+\Delta_2\nonumber\\
&+&\frac{1}{12}T^2\sum_i N_c h_{U_i}^2\left(\theta_{{\tilde U}_{L_i}}+
\theta_{{\tilde U}_{R_i}}+
\theta_{U_{L_i}}\theta_{U_{R_i}}
\right),\nonumber
\eear

\bear
\Pi_{H_2^0H_1^{0*}}&=&-\frac{1}{48}T^2\left[
(g_2^2+g_1^2)\theta_{H_2^0H_1^{0*}}
-2g_2^2 \theta_{H_2^{\pm}H_1^{\mp*}}    \right] +\Delta_{12},
\label{mixn}
\eear
\bear
\Pi_{H_2^\pm H_1^\mp}&=&-\frac{1}{48}T^2\left[
(g_2^2+g_1^2)\theta_{H_2^{\pm}H_1^{\mp*}}-2g_2^2
  \theta_{H_2^0H_1^{0*}}     \right] - \Delta_{12},
\label{mixc}
\eear
with
\bear
\Delta_1&=&\frac{T^2}{12}\sum_i\left\{N_c\left[h_{U_i}^2|\mu|^2
\theta_{{\tilde U}_{R_i}}^{{\tilde Q}_{L_i}} +
h_{D_i}^2 |A_{D_i }|^2
\theta_{{\tilde D}_{R_i}}^{{\tilde Q}_{L_i}}
\right]+ h_{E_i}^2 |A_{E_i}|^2
\theta_{{\tilde E}_{R_i}}^{{\tilde L}_i}
\right\},\,\,\nonumber\\
\Delta_2&=&\frac{T^2}{12}\sum_i\left\{N_c\left[h_{U_i}^2| 
A_{U_i}|^2
\theta_{{\tilde U}_{R_i}}^{{\tilde Q}_{L_i}} +
h_{D_i}^2 |\mu|^2
\theta_{{\tilde D}_{R_i}}^{{\tilde Q}_{L_i}}
\right]+ h_{E_i}^2 |\mu|^2
\theta_{{\tilde E}_{R_i}}^{{\tilde L}_i}
\right\},\,\,\nonumber\\
\Delta_{12}&=&\mu\frac{T^2}{12}\sum_i\left\{N_c\left[h_{U_i}^2 A_{U_i}
\theta_{{\tilde U}_{R_i}}^{{\tilde Q}_{L_i}}
+
h_{D_i}^2 A_{D_i}
\theta_{{\tilde D}_{R_i}}^{{\tilde Q}_{L_i}}
\right]+ h_{E_i}^2 A_{E_i}
\theta_{{\tilde E}_{R_i}}^{{\tilde L}_i}
\right\}.\nonumber
\eear

As an example of how to rotate $\theta$'s and $\Pi$'s consider the case 
in which only one (combination) of the Higgs  doublets is light compared 
with the temperature while the other is heavy and Boltzmann suppressed 
(this limit is realized for a large pseudoscalar mass and has been  
considered at finite temperature in  studies of the electroweak phase 
transition). We will concentrate in the Higgs loop contribution 
to Higgs thermal self-energies only. The rest of the terms are trivial 
to handle.
In terms of $\theta_H,\theta_\Phi$, the off-diagonal thermal mixing between 
${\overline { H}}_1$ and $H_2$ [eqs.~(\ref{mixn}),(\ref{mixc})] has the form
\bear
\Pi_{H_2^0H_1^{0*}}&=&-\frac{1}{48}T^2\left[
(g_2^2+g_1^2)(\theta_{H^0}-\theta_{\Phi^0})+2g_2^2(\theta_{H^\pm}
-\theta_{\Phi^\pm})\right]\sin\beta\cos\beta,\nonumber\\
\Pi_{H_2^\pm H_1^\pm}&=&\frac{1}{48}T^2\left[
(g_2^2+g_1^2)(\theta_{H^\pm}-\theta_{\Phi^\pm})+2g_2^2(\theta_{H^0}
-\theta_{\Phi^0})\right]\sin\beta\cos\beta.
\eear
In the 
neutral sector then, setting $\theta_{\Phi^0}=0$, it is easy to obtain
\be
\Pi_{H^0}=\Pi_{H_1^0}\cos^2\beta+\Pi_{H_2^0}\sin^2\beta+2
\Pi_{H_2^0H_1^{0*}}\cos\beta\sin\beta
=\frac{1}{48}(g_1^2+g_2^2)(2\theta_{H^0}
+\theta_{H^{\pm}})\cos^22\beta.
\ee
It can be checked that this is the correct result noting that the 
Standard Model result is 
\be
\Pi_H^{scalar}=\frac{1}{4}\lambda T^2,
\ee
(with the quartic Higgs coupling in the potential normalized 
to $V=\frac{1}{2}\lambda |H|^4$) while in the MSSM, the quartic self 
coupling of H, defined by (\ref{rotH}), is 
$\lambda=\frac{1}{4}(g_1^2+g_2^2)\cos^22\beta$.
\vspace{0.5cm}

{\bf D. GAUGE BOSONS}
\vspace{0.3cm}

{\bf D.1 $SU(3)_C$}
\bear
\Pi_{g_{L}}&=&\frac{1}{12}g_3^2T^2\left[4N_c\theta_g
+2\sum_{j}(\theta_{{\tilde U}_{L_{j}}}+
\theta_{{\tilde D}_{L_{j}}}+\theta_{{\tilde U}_{R_{j}}}+
\theta_{{\tilde D}_{R_{j}}})\right.\nonumber\\
&+&\left.\sum_{j}(\theta_{{U}_{L_{j}}}+
\theta_{{D}_{L_{j}}}+\theta_{{U}_{R_{j}}}+
\theta_{{D}_{R_{j}}})+2N_c\theta_{\tilde g}\right].
\eear

To simplify the contribution coming from squark loops we have used
\be
\theta_{\tilde q}+\theta_{\tilde q}^2=2\theta_{\tilde q}.
\label{simpl}
\ee
Note however that if we were to rotate the squark basis the expression on
the left hand side should be used.

{\bf D.2 $SU(2)_L$}
\bear
\Pi_{W_{3L}}&=&\frac{1}{24}g_2^2T^2\left[18\theta_{W^\pm}-2\theta_{gh}
+8\theta_{{\widetilde W}^\pm}
+N_c\sum_{j}(2\theta_{{\tilde U}_{L_{j}}}+
2\theta_{{\tilde D}_{L_{j}}}+
\theta_{{U}_{L_{j}}}+\theta_{{D}_{L_{j}}})\right.\nonumber\\
&+&
\sum_{j}(2\theta_{{\tilde \nu}_{L_{j}}}+2
\theta_{{\tilde 
e}_{L_{j}}}+\theta_{{\nu}_{L_{j}}}+\theta_{{e}_{L_{j}}})+2(
\theta_{H^0}+\theta_{H^\pm}+
\theta_{\Phi^0}+\theta_{\Phi^\pm})\\
&+&\left.\theta_{{\tilde H_1}^0}+\theta_{{\tilde 
H}_1^\pm} +  \theta_{{\tilde H}_2^\pm}+\theta_{{\tilde H}_2^0} 
\right]\nonumber,
\eear
\bear
\Pi_{W_{L}^\pm}&=&\frac{1}{24}g_2^2T^2\left[3(\theta_{W_3}+\theta_{W^\pm})
+12\theta_{W_3}\theta_{W^\pm}-2\theta_{gh}+
8\theta_{{\widetilde W}^\pm}\theta_{{\widetilde W}_3}\right.\nonumber\\
&+&N_c\sum_{j}\left[(\theta_{{\tilde U}_{L_{j}}}+
\theta_{{\tilde D}_{L_{j}}})^2+2
\theta_{{U}_{L_{j}}}\theta_{{D}_{L_{j}}}\right]+
\sum_{j}\left[(\theta_{{\tilde \nu}_{L_{j}}}+\theta_{{\tilde 
e}_{L_{j}}})^2+2\theta_{{\nu}_{L_{j}}}\theta_{{e}_{L_{j}}}\right]
\\
&+&
(\theta_{H^0}+\theta_{H^\pm})^2+(\theta_{\Phi^0}
+\theta_{\Phi^\pm})^2+\left.2\theta_{{\tilde H_1}^0}\theta_{{\tilde 
H}_1^\pm} +  2\theta_{{\tilde H}_2^\pm}\theta_{{\tilde H}_2^0} 
\right]\nonumber.
\eear

Here $\theta_{gh}$ gives the ghost piece and we have already rotated the 
Higgs contributions to the $H,\Phi$ basis using:
\be
\theta_{H_1^0}+\theta_{H_1^0}^2+\theta_{H_2^0}+\theta_{H_2^0}^2+
\theta_{H_2^0H_1^{0*}}^2=\theta_{H^0}+
\theta_{H^0}^2
\ee
and similarly for the charged $\theta$'s. We have simplify further
our expression using a relation similar to (\ref{simpl}).

{\bf D.3 $U(1)_Y$}
\bear
\Pi_{B_L}&=&\frac{1}{216}g_1^2T^2\left[
18\sum_j(\theta_{{\tilde \nu}_{L_j}}+\theta_{{\tilde e}_{L_j}}
+4\theta_{{\tilde e}_{R_j}})+
9\sum_j(\theta_{{\nu}_{L_j}}+\theta_{{e}_{L_j}}+4\theta_{{e}_{R_j}})
\right.\nonumber\\
&+&18(\theta_{H^\pm}+\theta_{H^0}+
\theta_{\Phi^\pm}+\theta_{\Phi^0})
+9(\theta_{{\tilde H}_1^\pm}+\theta_{{\tilde H}_1^0}+
\theta_{{\tilde H}_2^\pm}+\theta_{{\tilde H}_2^0})
\\
&+&\left.
2N_c\sum_j(\theta_{{\tilde U}_{L_j}}+\theta_{{\tilde 
D}_{L_j}}+16 \theta_{{\tilde U}_{R_j}}+4\theta_{{\tilde D}_{R_j}})
+N_c\sum_j(\theta_{{U}_{L_j}}+\theta_{{D}_{L_j}}
+16 \theta_{{U}_{R_j}}+4\theta_{{D}_{R_j}})\right].\nonumber
\eear
Here, contributions from scalars, and in particular Higgs bosons, have been
treated in the same way as explained for $SU(2)$ and $SU(3)$.

{\bf Acknowledgments}

We would like to thank M. Carena, M. Pietroni, M. Quir\'os and C.E.M. Wagner 
for useful discussions.

\end{document}